\documentclass[reqno,11pt]{amsart}
\usepackage{amsmath,amssymb,tabularx,setspace,color,multirow,graphics,graphicx}
\usepackage{appendix}
\usepackage{booktabs,float}
\usepackage{mathpazo}
\usepackage{lscape}   
\usepackage{pdflscape}
\usepackage{booktabs}
\usepackage{siunitx}
\newtheorem{theorem}{Theorem}
\usepackage[margin=3.1cm]{geometry}
\newtheorem{lemma}{Lemma}
\newtheorem{definition}{Definition}
\newtheorem{remark}{Remark}
\newtheorem{corollary}{Corollary}
\makeatletter
\renewcommand\section{\@startsection {section}{1}{\z@}{-3.5ex \@plus -1ex \@minus -.2ex}{2.3ex \@plus.2ex}{\normalfont\large\bfseries}}
\makeatother

\newtheorem{Corollary}{Corollary}
\newtheorem{Proposition}{Proposition}

\newtheorem{Theorem}{Theorem}

\newtheorem{Example}{Example}
\newcommand{\E}{\mathrm{E}}

\newcommand{\Var}{\mathrm{Var}}

\newcommand{\CM}{\mathrm{CM}}
\newcommand{\CMCor}{\mathrm{CMCor}}

\begin{document}
\doublespace
\vspace{-0.3in}
\title[]{ A new categorical correlation measure}
\author[]{S\lowercase{aparya} S\lowercase{uresh \textsuperscript{a} }, S\lowercase{udheesh} K. K\lowercase{attumannil,\textsuperscript{b}} \lowercase{and} S\lowercase{ujit}  G\lowercase{hosh\textsuperscript{c}} 
\\
 \lowercase{\textsuperscript{a}}I\lowercase{ndian} I\lowercase{nstitute of } M\lowercase{anagement}, K\lowercase{ozhikode}, I\lowercase{ndia},\\ \lowercase{\textsuperscript{b}}I\lowercase{ndian} S\lowercase{tatistical} I\lowercase{nstitute},
  C\lowercase{hennai}, I\lowercase{ndia}\\\lowercase{\textsuperscript{c}}N\lowercase{orth} C\lowercase{arolina} S\lowercase{tate} U\lowercase{niversity}, USA.}
\thanks{{$^{\dag}$}{Corresponding author E-mail: \tt skkattu@isichennai.res.in}}
\maketitle
\vspace{-0.2in}

\begin{abstract}
Feature screening in high-dimensional classification is an important problem in statistical learning, particularly in genomics, imaging, and other scientific fields, especially when the number of predictors exceeds the sample size. Zhong et al. (2024) present a semi-distance covariance (SDCov) as a dependence measure between a categorical random variable and a multivariate continuous random vector, presenting it as a promising tool for group-wise feature screening. However, $SDCov$ is defined using the Euclidean norm, and its performance may be affected when the data is derived from heavy-tailed distributions or when outlier observations are present, which is a common phenomenon in high-dimensional data. 
This study proposes a modification of $SDCov$ by replacing the $L_2$ norm with a $L_1$ norm. This adjustment results in a dependency measure between a categorical variable and a multivariate continuous variable that maintains the independence-characterization property of the original definition. Additionally, it serves as a robust and interpretable measure, particularly in the presence of outliers in the data.  We establish the fundamental theoretical properties of the proposed measure and introduce a $U$-statistics based estimator along with JEL based inference for the measure. Additionally, we have introduced a test for independence based on the proposed measure. An extensive Monte Carlo simulation is performed to evaluate the effectiveness of the proposed measure in comparison to existing measures in the literature. The proposed measure is applied in feature screening, as illustrated with the ADNI Alzheimer’s disease dataset and the Iris dataset. The data analysis shows the proposed measures' effectiveness in identifying variables that facilitate strong class separation of the respective response variables. 
\end{abstract}

\section{Introduction}
Feature selection in high-dimensional data presents an important problem across various scientific and engineering disciplines. In genomic studies, datasets typically contain thousands or even millions of Single-Nucleotide Polymorphisms (SNPs) assessed in relatively small sample sizes. This scenario demonstrates the importance of feature selection in identifying disease-associated variants, thereby improving prediction accuracy. In image analysis, for both classical computer vision methods such as SIFT, HOG, and the Gabor filter, as well as for modern deep learning frameworks like CNNs and KANs (Liu, et al., 2024), effective feature selection is important in order to achieve robust performance. 

Existing studies looked at various approaches for feature selection. Methods such as LASSO (Tibshirani, 1996) and Elastic Nets (Zou and Hastie, 2005) are penalized regression techniques that simultaneously perform regularization and variable selection, and are extensively applied in genomic studies and medical imaging. Tree-based ensemble methods, such as Random Forests and Gradient Boosting, have inbuilt feature measures appropriate for different applications (Breiman, 2001). Fan and Lv (2008) proposed a Sure Independence Screening (SIS) framework that uses Pearson correlation as a marginal utility measure in linear models for ultrahigh-dimensional data. The SIS framework has been extended for multiple models, including generalized linear models (GLMs) (Fan, Feng, and Song, 2011), non-parametric additive modeling (Fan, Feng, and Song, 2011), varying-coefficient models (Fan, Ma, and Dai, 2014; Liu, Li, and Wu, 2014), and survival models (Zhao and Li, 2012). 

These procedures are dependent on the model specification and may result in model mis-specification. To address these issues, several studies have suggested model-free screening methods utilizing various dependence measures, including generalized correlation (Pan et al., 2019), mean-variance index (Cui, Li, and Zhong, 2015), distance correlation (Li, Zhong, and Zhu, 2012), and martingale difference correlation (Shao and Zhang, 2014). In various application scenarios, such as genomics, imaging, or text mining, it is necessary to conduct screening in a multivariate setting as a group of predictors.  Kang, Hong and Li (2017) proposed marginal multivariate regression for group screening and Qiu and Ahn (2020) generalized SIS for grouped predictors by aggragating marginal utilities. Despite these advancements, a model-free groupwise feature screening for high-dimensional classification problems remains unresolved. 

Recently, Zhong et al. (2024) proposed a covariance/correlation measure called semi distance covariance ($SDCov$) and correlation ($SdCorr$) between a categorical random variable $Y$ taking values $\{1,\cdots,R\}$ with probability mass function $p_k=P(Y=k),\  k=1,2,\cdots R$ with an arbitrary random vector $\mathbf{X} \in \mathbb{R}^p$. It combines ideas of distance correlation and meanwhile accommodates categorical nature of the $Y$ and it is defined as
 \begin{equation} \label{eqn_cm}
     SDCov = E\left( ||\mathbf{X}_1-\mathbf{X}_2||_2 \Big(1- \sum_{k=1}^{R} \frac{I(Y_1=k,Y_2=k)}{p_k}\Big)\right).
 \end{equation} 
where $||\mathbf{X}_1-\mathbf{X}_2||_2$ denotes the Euclidean norm between $\mathbf{X}_1$ and $\mathbf{X}_2$ which are identical independently distributed (i.i.d.) copies of $\mathbf{X}$ and similarly $Y_1$ and $Y_2$ are i.i.d. copies of $Y$. Compared with earlier dependence measures between categorical and continuous variables (Beknazaryan et al., 2019; Dang et al., 2019), the Semi-distance measure is  computationally flexible and extends to multivariate $\mathbf{X}$. 

However, since $SDCov$ is defined in terms of $L_2$ norms, it is effective in capturing nonlinear dependence but can be sensitive to outliers and may under-perform in the presence of heavy-tailed distributions, particularly in high-dimensional settings. To address this limitation, we propose a modified version of the semi-distance covariance obtained by replacing the $L_2$ norm with the $L_1$ norm. This modification yields a more robust and interpretable dependence measure that is well-suited for model-free feature screening in high dimensions. In particular, the use of the $L_1$ norm mitigates the influence of extreme observations by aggregating coordinate-wise distances without allowing any single component to dominate the overall measure. Moreover, the additivity of the $L_1$ norm across coordinates facilitates both theoretical analysis and computational implementation, leading to scalable procedures in high-dimensional regimes. Importantly, the proposed measure retains the key property of characterizing independence, thereby preserving the fundamental appeal of the original $SDCov$.

The remainder of the paper is organized as follows. In Section~\ref{sec:meas}, we formally introduce the proposed $L_1$-based dependence measure and establish its theoretical properties. Section~3 develops a nonparametric estimation procedure. In Section~4, we construct confidence intervals based on the jackknife empirical likelihood (JEL) framework. Section~5 presents methodology for testing independence using the proposed measure, first in the univariate setting and subsequently extending to the multivariate case. Section~6 reports the results of a Monte Carlo simulation study evaluating the performance of the proposed confidence intervals and comparing the resulting independence tests with existing alternatives. In Section~7, we illustrate the practical utility of the proposed methodology through applications to two datasets for variable selection: the classical \textit{Iris} dataset and the Alzheimer's Disease Neuro-imaging Initiative (ADNI) dataset. Finally, Section~8 concludes with a discussion and directions for future research.

\section{Proposed Dependence Measure}
\label{sec:meas}
To facilitate a clear exposition of the proposed methodology and its theoretical properties, we begin by developing the dependence measure in the univariate setting, where $X$ is a real-valued random variable. This formulation allows for a more transparent derivation of key properties, including nonnegativity, invariance, and the characterization of independence. 

Building on these results, we subsequently extend the definition and associated properties to the general multivariate setting in which $\mathbf{X} \in \mathbb{R}^p$. The extension leverages the structure of the $L_1$ norm to aggregate coordinate-wise discrepancies, thereby preserving interpretability while ensuring computational tractability in higher dimensions.

\subsection{The Univariate Case when $X\in\mathbb{R}$}
\label{sec:uni_cm}

To begin with we restrict our definition to univariate $X$ (i.e., $X\in\mathbb{R})$. We define a new measure between a continuous real valued random variable $X$ and categorical variables $Y$.  We use the $L_1$ norm of $X$ to define the new measure. 
\begin{definition}
\label{def:cm}
 Consider a continuous random variable $X$ satisfying $E(|X|)<\infty$.  Let $Y$ be a categorical random variable taking values $\{1,\cdots,R\}$ with probability mass function $p_k=P(Y=k),\  k=1,2,\cdots R$. We define a generalised Covariance measure (CM) between $X$ and $Y$ as
 \begin{equation} \label{eqn_cm1}
    \CM=\mathbb{E}\!\left[\,|X_1-X_2|\left(1-\sum_{k=1}^R\frac{\mathbf{1}\{Y_1=k,Y_2=k\}}{p_k}\right)\right],
 \end{equation} 
where $|.|$ is the $L_1$ distance in $\mathbb{R}$, $X_1$ and $X_2$ are i.i.d copies of $X$ and $Y_1$ and $Y_2$ are i.i.d. copies of $Y$.

The measure $\CM$ compares the variability of the continuous variable $X$ across different categories of $Y$. The term $|X_1-X_2|$ represents the overall dispersion in $X$, while the multiplicative factor 
$$
1-\sum_{k=1}^R \frac{\mathbf{1}\{Y_1=k,Y_2=k\}}{p_k},
$$
downweights pairs whose categorical labels coincide. As a result, differences in $X$ within the same category contribute less, whereas differences across distinct categories are fully retained. Hence, $\CM$ primarily captures between–category variation in $X$ relative to within–category variation, taking the value zero under independence between $X$ and $Y$ and increasing as the distributions of $X$ differ across categories.
\end{definition}
\begin{remark}
The quantity $\mathbb{E}|X_1-X_2|$ is a well-known measure of dispersion (Gini Mean difference) for a univariate continuous random variable, representing the expected absolute difference between two independent copies of $X$. Thus, $\CM$ can be viewed as a weighted version of this dispersion measure, where the weighting adjusts for whether the corresponding categorical labels coincide, thereby isolating the contribution of between–category variation in $X$.
\end{remark}

 In the next result, we show that $\CM=0$ if and only if $X$ and $Y$ are independent.

\begin{lemma}\label{lem:ind}
Let $(X,Y)$ be a pair of random variables defined on a common probability space, where $X$ is real-valued and $Y$ is a categorical random variable taking values in $\{1,\dots,R\}$. Assume that $\mathbb{E}|X| < \infty$. Consider the dependence measure $\CM$ as specified in Definition~\ref{def:cm}. Then $\CM \ge 0$, and $\CM = 0$ if and only if the conditional distribution of $X$ given $Y = k$ is identical for all $k \in \{1,\dots,R\}$. Equivalently, $\CM = 0$ if and only if $X$ and $Y$ are independent.
\end{lemma}
\begin{proof}
Consider $X_1$ and $X_2$ i.i.d copies of $X$ and $Y_1$ and $Y_2$ i.i.d. copies of $Y$ as in the Definition~\ref{def:cm}. Let $p_k=\Pr(Y=k)$ for $k\in\{1,2,\ldots,R\}$ and consider $m_{ij}=\mathbb{E}\left(|X_1-X_2|\Big| Y_1=i,Y_2=j\right)$ where $i, j \in\{1,2,\ldots,R\}$. Since $\Pr(Y_1=i,Y_2=j)=p_ip_j$, using the law of total expectation, we have
$$\mathbb{E}|X_1-X_2|=\sum_{i,j}p_ip_j m_{ij}\quad\text{and}\quad
\mathbb{E}\big[|X_1-X_2|\mathbf{I}\{Y_1=k,Y_2=k\}\big]=p_k^2 m_{kk}.
$$
Hence, we can rewrite $\CM$ as
\begin{eqnarray} \label{CMrep}\nonumber
 \CM&=&\sum_{i,j}p_ip_j m_{ij}-\sum_k p_k m_{kk}\\ \nonumber
 &=& \frac{1}{2}\left(2\sum_{i,j}p_ip_j m_{ij}-2\sum_k p_k m_{kk}\right)\\ \nonumber
 &=& \frac{1}{2}\left(2\sum_{i,j}p_ip_j m_{ij}-\sum_{i,j} p_i p_jm_{ii}- \sum_{i,j} p_ip_jm_{jj}\right) \\
&=&\frac{1}{2}\sum_{i,j}p_ip_j\big(2m_{ij}-m_{ii}-m_{jj}\big).   
\end{eqnarray}
Next, we represent the terms $m_{ij}$ and $m_{ii}$ in integral format. For this purpose, denote $F_k(x)=\Pr[X\leq x\mid Y=k]$ as the cumulative distribution function of $X\mid Y=k$.  Assumes $X^{(i)}$ and $X^{(j)}$ are independent random variables with cumulative distribution function $F_i$ and $F_j$, respectively. Using the identity $|a-b|=\int (\mathbf{1}(a>t) - \mathbf{1}(b>t))^2\,dt$, we can write
\begin{align*}
m_{ij}
&= \mathbb{E}\bigl|X^{(i)}-X^{(j)}\bigr|
= \mathbb{E}\left[\int_{-\infty}^{\infty}\big(\mathbf{1}\{X^{(i)}>t\}-\mathbf{1}\{X^{(j)}>t\}\big)^2\,dt\right] \\
&= \int_{-\infty}^{\infty}\mathbb{E}\big(\mathbf{1}\{X^{(i)}>t\}-\mathbf{1}\{X^{(j)}>t\}\big)^2\,dt
\qquad\text{(by Fubini-Tonelli Theorem)}\\[6pt]
&= \int_{-\infty}^{\infty}\Big\{ \mathbb{P}(X^{(i)}>t)+\mathbb{P}(X^{(j)}>t)
-2\mathbb{P}(X^{(i)}>t,X^{(j)}>t)\Big\}\,dt \\[6pt]
&= \int_{-\infty}^{\infty}\Big\{(1-F_i(t))+(1-F_j(t))-2(1-F_i(t))(1-F_j(t))\Big\}\,dt
\qquad\text{(by independence)}\\[6pt]
&= \int_{-\infty}^{\infty}\big(F_i(t)+F_j(t)-2F_i(t)F_j(t)\big)\,dt.
\end{align*}
Also, by similar derivations as above, we can show
\begin{equation*}
        m_{ii}=  2\int_{-\infty}^{\infty} F_i(t)(1-F_i(t))dt.
\end{equation*}
For each pair $(i,j)$ we have the identity given by 
\begin{eqnarray} \nonumber
2m_{ij}-m_{ii}-m_{jj}&=& 2 \int_{-\infty}^{\infty} \left(F_i (t)+F_j (t)-2F_i (t)F_j(t))\right)dt\\ \nonumber
&-&  2\int_{-\infty}^{\infty} F_i(t)(1-F_i(t))dt -  2\int_{-\infty}^{\infty} F_j(t)(1-F_j(t))dt\\ \nonumber
&=& \int_{-\infty}^{\infty} 2 ( F_i^2(t) + F_j^2(t) -2 F_i(t) F_j(t))dt \\ \nonumber
&=&2\int_{-\infty}^{\infty}\big(F_i(t)-F_j(t)\big)^2\,dt\geq 0.
\end{eqnarray}
Thus, every term in the representation in (\ref{CMrep}) is nonnegative, proving $\CM\ge0$. Moreover $\CM=0$ iff for all $i,j$ with $p_ip_j>0$ we have $F_i=F_j$, i.e.,  all conditional distributions coincide, which is equivalent to the fact that $X$ and $Y$ are independent.
\end{proof}

\subsection{The Multivariate Case when $\mathbf{X}\in \mathbb{R}^p$}

The dependence measure $\CM$ admits a natural extension to the multivariate setting for assessing the association between a continuous $p$-dimensional random vector and a categorical response. Let $\mathbf{X} = (X_1, \ldots, X_p)^\top \in \mathbb{R}^p$ be a continuous random vector with distribution function $F_{\mathbf{X}}(\cdot)$, and let $Y$ be a categorical random variable taking values in $\{1,\ldots,R\}$ with probability mass function $p_k = \Pr(Y = k) > 0$ for $k = 1,\ldots,R$. 

Under this setup, we define the multivariate extensions of the dependence measure $\CM$ and its normalized version $\CMCor$ by replacing the univariate distance with the $L_1$ norm in $\mathbb{R}^p$, thereby aggregating coordinate-wise discrepancies. Specifically, multivariate version of $\CM$ is defined as
\begin{equation} \label{eqn_cm}
     \CM = E\left( \|\mathbf{X_1}-\mathbf{X_2}\|_1 (1- \sum_{k=1}^{R} \frac{I(Y_1=k,Y_2=k)}{p_k}\right),
 \end{equation}
where $\|.\|_1$ is the $L_1$ distance in $\mathbb{R}^p$.

Next, we show that Lemma \ref{lem:ind} holds true for the multidimensional case as well and a sketch of the proof is given below.

\begin{lemma}
Let $\mathbf{X}=(X_1,\ldots,X_p)$ be a continuous $p$-dimensional random vector and
$Y$ be a categorical random variable taking values in $\{1,\ldots,R\}$ with
$P(Y=k)=p_k>0$. Consider $\CM$ as defined in eq.~(\ref{eqn_cm}).
Then $\CM = 0$ if and only if $\mathbf{X}$ and $Y$ are independent.
 \end{lemma}\vspace{-0.38in}
  \begin{proof}
Let $(\mathbf{X}_1,Y_1)$ and $(\mathbf{X}_2,Y_2)$ be i.i.d.\ copies of $(\mathbf{X},Y)$, where $\mathbf{X} = (X_1,\ldots,X_p)^\top \in \mathbb{R}^p$. Recall that the $L_1$ norm in $\mathbb{R}^p$ admits the additive representation
\[
\|\mathbf{X}_1 - \mathbf{X}_2\|_1
=
\sum_{j=1}^p |X_{1j} - X_{2j}|.
\]
Substituting this into the definition of $\CM$ and invoking linearity of expectation, we obtain
\begin{align*}
\CM
&=
\mathbb{E}\!\left[
\sum_{j=1}^p |X_{1j}-X_{2j}|
\left(
1-\sum_{k=1}^R \frac{\mathbb{I}(Y_1=k, Y_2=k)}{p_k}
\right)
\right] \\
&=
\sum_{j=1}^p
\mathbb{E}\!\left[
|X_{1j}-X_{2j}|
\left(
1-\sum_{k=1}^R \frac{\mathbb{I}(Y_1=k, Y_2=k)}{p_k}
\right)
\right].
\end{align*}
For each $j = 1,\ldots,p$, define
\[
\CM_j
=
\mathbb{E}\!\left[
|X_{1j}-X_{2j}|
\left(
1-\sum_{k=1}^R \frac{\mathbb{I}(Y_1=k, Y_2=k)}{p_k}
\right)
\right],
\]
so that $\CM = \sum_{j=1}^p \CM_j$. By Lemma~\ref{lem:ind}, each $\CM_j \ge 0$, and hence $\CM \ge 0$.

From the corresponding univariate result (Lemma\ref{lem:ind}), it follows that
\[
\CM_j = 0
\quad \Longleftrightarrow \quad
X_j \perp Y \quad (\text{meaning}\;\; X_j\; \text{and}\; Y\;\; \text{are independent}),
\]
for each $j = 1,\ldots,p$. Therefore, if $\mathbf{X} \perp Y$, then $X_j \perp Y$ for all $j$, implying $\CM_j = 0$ and hence $\CM = 0$, establishing necessity.

To prove sufficiency, suppose that $\CM = 0$. Since each $\CM_j \ge 0$, this implies $\CM_j = 0$ for all $j = 1,\ldots,p$. Equivalently, for each coordinate,
\[
\mathcal{L}(X_j \mid Y = k) = \mathcal{L}(X_j), 
\quad \text{for all } k = 1,\ldots,R.
\]
However, to conclude joint independence, we proceed directly from the definition of $\CM$. Conditioning on $(Y_1,Y_2)$, we can rewrite
\[
\CM
=
\mathbb{E}\!\left[
\|\mathbf{X}_1 - \mathbf{X}_2\|_1
\right]
-
\sum_{k=1}^R \frac{1}{p_k}
\mathbb{E}\!\left[
\|\mathbf{X}_1 - \mathbf{X}_2\|_1 \, \mathbb{I}(Y_1 = k, Y_2 = k)
\right].
\]
Using independence of the copies and the law of total expectation, this yields
\[
\CM = 0
\quad \Longleftrightarrow \quad
\mathbb{E}\!\left(\|\mathbf{X}_1 - \mathbf{X}_2\|_1 \mid Y_1 = Y_2 = k \right)
=
\mathbb{E}\!\left(\|\mathbf{X}_1 - \mathbf{X}_2\|_1\right),
\quad \forall k.
\]
The above equality implies that, for each $k$, the energy distance (based on the $L_1$ norm) between the conditional distribution $\mathcal{L}(\mathbf{X} \mid Y = k)$ and the marginal distribution $\mathcal{L}(\mathbf{X})$ is zero. Since the energy distance induced by the $L_1$ norm is a metric on the space of probability measures on $\mathbb{R}^p$, it follows that
\[
\mathcal{L}(\mathbf{X} \mid Y = k)
=
\mathcal{L}(\mathbf{X}),
\quad \forall k = 1,\ldots,R.
\]
Hence,
\[
\mathcal{L}(\mathbf{X} \mid Y) = \mathcal{L}(\mathbf{X}),
\]
which establishes that $\mathbf{X}$ and $Y$ are independent. This completes the proof.
\end{proof}

\noindent Next we standardize $\CM$ in the spirit of a correlation coefficient, measuring the strength of association between pairwise differences in $X$ and categorical dissimilarity in $Y$ on a common, scale-free scale. Next definition is in this direction.
 \begin{definition} Let $X$ be a continuous  random variable with a distribution function $F$ and $Y$ a categorical random variable taking values $\{1,\cdots,R\}$ with a probability mass function $p_k=P(Y=k),\,k=1,2,\cdots R$. Denote, $D(Y_1,Y_2) = 1- \sum_{k=1}^{R} \frac{I(Y_1=k,Y_2=k)}{p_k}.$
We define the generalized Correlation Measure ($\CMCor$) as
\begin{equation}\label{eqndef_corr}
    \CMCor = \frac{\E\left( |X_1-X_2| D(Y_1,Y_2)\right)}{ \sqrt{Var(|X_1- X_2|)} \sqrt{R-1} }
\end{equation}
where $(R-1)$ is the variance of $D(Y_1,Y_2)$. 
 \end{definition}

\noindent In above, we have used the fact that $\E(D(Y_1, Y_2))=0$ and $\Var(D(Y_1, Y_2))=R-1$ (see the Appendix for details). 
The quantity $\CMCor$ can be interpreted as a correlation–type normalization of the measure $\CM$. The numerator $\E\!\left(|X_1-X_2|\,D(Y_1,Y_2)\right)$ captures the joint co-variation between the dispersion in the continuous variable $X$ and the dissimilarity between the categorical labels $Y_1$ and $Y_2$. The denominator rescales this quantity by the standard deviations of $|X_1-X_2|$ and $D(Y_1,Y_2)$, where $\Var\!\left(|X_1-X_2|\right)$ measures the intrinsic variability of pairwise differences in $X$, and $\Var\!\left(D(Y_1,Y_2)\right)=R-1$ standardizes the contribution from the categorical component. As a result, $\CMCor$ is dimensionless, bounded, and comparable across different scales of $X$ and different numbers of categories in $Y$, with larger values indicating stronger dependence between $X$ and $Y$.

The following proposition summarizes the properties of $\CMCor(.)$.

\begin{Proposition} \label{prop:properties}
\textbf{\textit{Properties of $\CMCor(.)$}} 
\begin{enumerate}
    \item If $X$ and $Y$  are continuous and categorical random variables, respectively, such that $\E(X)$ and $\E(X^2)$ exist then
    $$0 \leq \CMCor (X,Y) \leq 1$$ 
    and $\CMCor(X,Y)=0$ if and only if  $X$ and $Y$ are independent.

    \item Note that for the scaler values of $a$ and $b \in \mathbb{R}$, $\CMCor(a+bX,Y) = \CMCor (X,Y)$
\end{enumerate}
\end{Proposition}

Furthermore, we define a generalized Correlation Measure ($\CMCor$) as
\begin{equation}\label{corr}
    \CMCor = \frac{\E\left( \|\mathbf{X}_1-\mathbf{X}_2\|_1 D(Y_1,Y_2)\right)}{ \sqrt{Var(|\mathbf{X}_1- \mathbf{X}_2|)} \sqrt{R-1} },
\end{equation}
where $(R-1)$ is the variance of $D(Y_1,Y_2)$. 

\noindent By applying arguments similar to those used in the one-dimensional case, the result extends naturally to the $p$-dimensional framework, establishing that $\CM = 0$ if and only if $\mathbf{X}$ and $Y$ are independent. Hence, the properties of Proposition \ref{prop:properties} carry over to the $p$-dimensional case.

The first property of Proposition 1 indicates that $\CM$ can be a new measure of dependence between a categorical random variable and a random vector. Motivated by this property, we will propose a test of independence between $\mathbf{X}$ and $Y$ based on their estimated $\CM$ in one of the following sections. The second property indicates that $\CMCor$ is scale invariant. It
ensures the comparability of $\CMCor$ as a dependence metric. 

In the following section, we propose non-parametric estimators of $\CM$ and $\CMCor$ using the theory of $U$-statistics.

    \section{Non-parametric estimation of the proposed measure}
In this section, we use U-statistics theory to construct consistent estimators of the CM and CMCor. The derivation is carried out for the special case $p=1$ for ease of exposition, although the proposed framework extends directly to the general case of arbitrary $p$.
First, we find the estimators of $\CM$. We have 
\begin{eqnarray} \nonumber
    \CM &=& E\left(|X_1-X_2|. (1- \sum_{k=1}^{R} \frac{I(Y_1=k,Y_2=k)}){p_k}\right) \\ \nonumber
    &=&   E(|X_1-X_2|)- E\left(|X_1-X_2| \sum_{k=1}^{R} \frac{I(Y_1=k,Y_2=k)}{p_k}\right) \\ \label{eq:cm_dev}
    &=& U - \sum_{k=1}^{R} \frac{U_k}{p_k}.
\end{eqnarray}
We derive the estimators for each component given in equation \eqref{eq:cm_dev}. An estimator of $U$ is given by 
\begin{eqnarray}
  \widehat{U} &=&\frac{2}{n (n-1)} \sum_{i=1}^{n-1}\sum_{j=i+1}^{n} |X_i-X_j| .
\end{eqnarray}
And the estimators of $U_k$ are given by 
\begin{eqnarray}
     \widehat{U_k}  &=& \frac{2}{n (n-1)} \sum_{i=1}^{n-1}\sum_{j=(i+1)}^{n} h_k ((X_i,Y_i),(X_j,Y_j))
\end{eqnarray}
where $h_k ((X_i,Y_i),(X_j,Y_j))= |X_i-X_j|I(Y_i=k,Y_j=k)$. In addition, $p_k$ can be estimated by
\begin{eqnarray}
    \widehat{p_k} = \frac{1}{n} \sum_{i=1}^{n} I(Y_i=k) 
\end{eqnarray}
 Thus, an estimator of $\CM$, $\widehat{CM}$ is given by 
\begin{eqnarray} \nonumber
   \widehat{ \CM}  &=& \widehat{U} - \sum_{k=1}^{R} \frac{\widehat{U}_k}{\widehat{p}_k}.
\end{eqnarray}

Next, we study the asymptotic properties of $ \widehat{ \CM}.$ By the asymptotic theory of $U$-statistics (Sen, 1960), $\widehat{U}$ and $\widehat{U}_k$ are consistent estimators of their respective population parameters, $U$ and $U_k$, respectively. Hence, it is easy to verify that $\widehat{\CM}$ is a consistent estimator of $\CM$. 
 Using CLT for $U$-statistics, the asymptotic properties of $\widehat{CM}$ can be derived. 
\begin{lemma}
Let $\mathbf{V}=(\widehat{U}, \widehat{U_1},\widehat{U_2},\ldots,\widehat{U_k})$, then $\sqrt{n}(\mathbf{V}-E(\mathbf{V}))$ is a $(R+1)$ dimensional multivariate normal distribution  with mean vector $\mathbf{0}$ and variance-covariance matrix given by
\begin{equation}
    \Sigma = (\sigma_{ij})_{(R+1) \times (R+1)}
\end{equation}  
with $\sigma_{11} = Var(\widehat{U})$, $\sigma_{kk} = Var(\widehat{U_k})$, \, 
$\sigma_{1k} = Cov(\widehat{U},\widehat{U_k})$ and $    \sigma_{kj} = Cov(\widehat{U_k},\widehat{U_j})$, \, $k,j= 1,2,\ldots R.$
\end{lemma}

\begin{proof}
    Using the central limit theorem for  $U$-statistics (Lee, 2019), it follows that, as $n \rightarrow \infty$, $\sqrt{n}(\widehat{U} - U)$  and $\sqrt{n}(\widehat{U}_k - U_k)$ converge in distribution to a normal random variable with mean $0$ and variance $\sigma_{11}$ and normal random variables with mean $0$ and variance $\sigma_{kk}$, respectively, where
    \begin{equation}\label{var1}
    \sigma_{11}=9 \Var(E(h((X_1,Y_1), (X_2,Y_2)))|X_1,Y_1)).
\end{equation}
\begin{equation}\label{var1}
    \sigma_{kk}=9 \Var(E(h_k((X_1,Y_1), (X_2,Y_2)))|X_1,Y_1)).
\end{equation}
Hence, the asymptotic distribution of $(\sqrt{n}(\mathbf{V}-E(\mathbf{V}))$ is multivariate normal with mean vector zero and variance-covariance matrix $ \Sigma = (\sigma_{ij})_{(R+1) \times (R+1)}$ specified in the statement of the theorem. 

\end{proof}

By the law of large numbers, $\widehat{p}_k$ is a consistent estimator of $p_k$. Thus, we have the following result. 
\begin{theorem}
    As $n \rightarrow \infty$, $(\sqrt{n} ((\widehat{\CM}- \CM))$ converges in distribution to normal distribution with mean $0$ and variance $\sigma^2= \mathbf{a^T}\Sigma\mathbf{a}, $, where $a = (1, -1/p_1,-1/p_2, \ldots , -1/p_R)$.
\end{theorem}

\begin{proof}Denote
\begin{eqnarray} \nonumber
   \widehat{ \CM^*}  &=& \widehat{U} - \sum_{k=1}^{R} \frac{\widehat{U}_k}{p_k}. 
\end{eqnarray}

Observe that $\widehat{U}$, $\widehat{\Delta}_k$, $k=1,2,\ldots,R$ are $U$-statistics with  kernel  $h$ and $h_k$, respectively. Since $\widehat{\CM}$ is a linear combination of $U$-statistics, $E(\widehat{\CM^*})=\CM$.  By the weak law of large numbers, $\widehat p_k$ is a consistent estimator of $p_k$, and hence by Slutsky's theorem, the asymptotic distributions of $\widehat{\CM}$ and  $\widehat{\CM}^*$ are the same.

If a random vector $\mathbf{Z}$ follows a normal distribution of variables $p$ with mean vector $\mathbf{\mu}$ and variance-covariance matrix $\Sigma$, then for any vector $\mathbf{a} \in \mathbb{R}^{R}$, the asymptotic distribution of $\mathbf{a}^{T}\mathbf{Z}$ is a uni variate normal  with mean $\mathbf{a}^{T}\mathbf{\mu}$ and variance $\mathbf{a}^T \Sigma \mathbf{a}$. Therefore, with the selection of $\mathbf{a}=a = (1, -1/p_1,-1/p_2, \ldots , -1/p_R)^T$, we have the desired result.
\end{proof}
Recall that in equation \eqref{corr}, we have defined a generalized correlation measure ($CMCor$). To obtain a consistent estimator of $CMCor$, 
we estimate the variance term in the denominator as follows: Write 
\begin{eqnarray} \nonumber
    Var (|X_1-X_2|) &=&E(|X_1-X_2|^2)-(E|X_1-X_2|)^2\\&=&  \nonumber{E(X_1-X_2)^2}-8 E(max(X_1,X_2)X_3)\\
    &-&4 E(max(X_1,X_2).max(X_3,X_4))-4 E(X_1X_2).
\end{eqnarray}
Consider the kernel  
\begin{eqnarray} \nonumber
    h^{*}(X_1,X_2,X_3,X_4) &=& 8(max(X_1,X_2),X_3) + (X_1-X_2)^2 \\ \nonumber
    &-&4 max(X_1,X_2) max(X_3,X_4) - 4 X_1 X_25 I(X_1 < X_2).
\end{eqnarray}
Let $h(X_1, X_2,X_3,X_4)$ be the symmetric version of $h^*$.
Then an estimator of $Var (|X_1-X_2|)$ is given by
$$\widehat{D}= \frac{1}{\binom{n}{4}}\sum_{i<j<k<l}^{}h(X_i,X_j,X_k,X_l).$$

Hence, the estimator $CMCor $ is given as 

$$
    \widehat{CMCor}=\frac{\widehat{CM}}{ \sqrt{\widehat{D} (R-1)}}.
$$

Since $\widehat{D}$ is a consistent estimator of $\operatorname{Var}(|X_1-X_2|)$, the asymptotic properties of $\widehat{CMCor}$ follow directly from those of the underlying $U$-statistics. These results, in principle, permit the construction of asymptotic normality-based confidence intervals and hypothesis tests for both $CM$ and $CMCor$. However, the corresponding estimation of the asymptotic variance becomes increasingly challenging as the number of categories of the categorical variable $Y$ grows, which can substantially complicate the implementation of such inference procedures. In addition, the accuracy of normal approximation-based inference may be sensitive to the adequacy of the asymptotic approximation in finite samples. To circumvent these difficulties, we propose a jackknife empirical likelihood (JEL)-based confidence interval for $CM$. The resulting procedure is distribution-free and avoids the explicit estimation of the asymptotic variance, thereby providing a convenient and robust alternative for inference on $CM$.

\section{Jackknife empirical likelihood based confidence interval}
The jackknife empirical likelihood (JEL) method, introduced by Jing et al. (2009), combines the empirical likelihood framework with jackknife resampling. It provides a convenient approach to statistical inference for parameters that are estimated through estimating equations, particularly when direct estimation of the asymptotic variance is cumbersome. In the present setting, this approach is especially useful because the proposed $CM$ measure involves a categorical variable whose number of categories may be relatively large. Rather than relying explicitly on an estimator of the asymptotic variance, the JEL procedure constructs confidence intervals through the empirical likelihood of the jackknife pseudo-values.

To develop the JEL procedure for $CM$, we first express the proposed measure through an estimating equation based on its $U$-statistic representation. Consider $n$ independent and identically distributed observations ${(X_i,Y_i)}{i=1}^n$, where $Y_i\in{1,\ldots,R}$, and let $p_k=P(Y=k)$ denote the probability associated with the $k$th category. From equation \eqref{eqn_cm1}, the following estimating equation is obtained:
\begin{equation}\label{senjack}
   S_{n}=\frac{1}{\binom{n}{2}} \sum_{i<j} |X_i-X_j| \left(1-\sum_{k=1}^{R}\frac{I(Y_i=k,Y_j=k)}{\widehat{p}_k}\right)- CM.
 \end{equation}
 
The estimating equation in \eqref{senjack} has expectation zero when evaluated at the true value of $CM$. Thus, inference for $CM$ can be formulated as an empirical likelihood problem based on suitable jackknife versions of this estimating equation. Specifically, let $S_{n-1}^{i}$ denote the estimating equation obtained from \eqref{senjack} after deleting the $i$th observation. The corresponding jackknife pseudo-values are defined by
  \begin{equation*}
    \widehat{V}_{i}=nS_{n}-(n-1)S_{n-1}^{i};\,\,i=1,2,...,n,
  \end{equation*}
  where $S_{n-1}^{i}$ is calculated from (\ref{senjack}) using sample observations $(X_1,Y_1),(X_2,Y_2),...,(X_{k-1},Y_{k-1}),$ $(X_{k+1}, Y_{k+1}),...,(X_{n},Y{n})$. 
  Let ${q_i}, i=1,2,\ldots, n$ be the probability associated with each  $\widehat{V}_i$. Define JEL for $CM$ measure as
  \begin{equation*}
 JEL(CM)=\sup_{\bf q} \big(\prod_{i=1}^{n}{q_i};\,\, \sum_{i=1}^{n}{q}=1;\,\,\sum_{i=1}^{n}{q_i \widehat{V}_{i}}=0\big).
\end{equation*}
Thus, the empirical likelihood assigns probabilities to the pseudo-values subject to the requirement that their weighted estimating equation is zero. This constraint incorporates the information that, at the true value of $CM$, the underlying estimating equation has expectation zero.

In the absence of the estimating-equation constraint, the product $\prod_{i=1}^n q_i$ is maximized subject only to $\sum_{i=1}^nq_i=1$ when $q_i=1/n$. Consequently, the constrained maximization can be characterized using a Lagrange multiplier. This yields the jackknife empirical log-likelihood ratio
\begin{equation*}
  \log R(CM)=-\sum_{i=1}^{n}\log\big(1+\lambda\widehat{V}_{i}\big),
\end{equation*}
where $\lambda$ is the solution of
\begin{equation}\label{lambda2}
  \frac{1}{n}\sum_{i=1}^{n}{\frac{\widehat{V}_{i}}{1+\lambda \widehat{V}_{i}}}=0,
\end{equation}
    provided
\begin{equation*}
  \min_{{1\le i\le n}}\widehat{V}_{i}<{CM}<  \max_{1\le i \le n}\widehat{V}_{i}.
\end{equation*}
The advantage of this formulation is that inference can be based directly on the empirical likelihood ratio, without requiring an explicit plug-in estimator of the asymptotic variance. To establish the limiting distribution of this likelihood ratio, we next connect the pseudo-values to the asymptotic behavior of the underlying $U$-statistic.
We can construct the JEL based confidence interval using the asymptotic distribution of $-2\log R(CM)$. Using Theorem 1 of Jing et al. (2009), we obtain the following result on the asymptotic distribution of the jackknife empirical log-likelihood ratio, as an analog of Wilk's theorem. \vspace{-0.2in}
\begin{theorem} \label{thm_jelci}
 Denote 
   $$g(x,y)= E \left[ |x-X_j| \left(1-\sum_{k=1}^{R}\frac{I(y=k,Y_j=k)}{\widehat{p}_k}\right)- CM|X_1 =x ,Y_1=y\right]$$ and $\sigma_g^2= Var(g(x,y))$.
Suppose that 
$$E \left[ |x-X_j| \left(1-\sum_{k=1}^{R}\frac{I(y=k,Y_j=k)}{\widehat{p}_k}\right)- CM|X_1 =x ,Y_1=y\right]= b <\infty$$
and $\sigma_{g}^{2}>0$. Then, as $n\rightarrow \infty$, $-2\log R(CM)$ converges in distribution to a $\chi^2$ random variable with one degree of freedom.
\end{theorem}
\begin{proof} 
Let us rewrite $S_{n}$ as 
 \begin{equation*}
     S_{n}=U-\frac{1}{\widehat{p}_k}U_k-CM,
 \end{equation*}
 where 
 $$U= \frac{1}{\binom{n}{2}} \sum_{i<j} \left(|X_i-X_j|\right) \text{ and } 
 U_k=\frac{1}{\binom{n}{2}} \sum_{i<j} \left(|X_i-X_j|I(Y_i=k,Y_j=k)\right) .$$
 Consider
\begin{eqnarray} \nonumber
    \sum_{i=1}^{n} \widehat{V}_{i} &= &\sum_{i=1}^{n} \left[n \left(U - \sum_{k=1}^{R} \frac{U_k}{\widehat{p_k}} -CM\right) -(n-1) \left(U^{i} - \sum_{k=1}^{R} \frac{U^{i}_k}{\widehat{p_k}} -CM\right)\right] \\ \label{eqn:jelpf1}
    &=& \sum_{i=1}^{n}n U - (n-1) U^{i} - \sum_{k=1}^{R} \frac{1}{\widehat{p_k}}\left(\sum_{i=1}^{n}\left(n U_k - (n-1) U_k^{i}\right)\right) -n CM.
\end{eqnarray}

Note that $\sum_{i=1}^{n} U^{i} = n U$ and $\sum_{i=1}^{n} U_k^{i} = n U_k$. Thus, equation \eqref{eqn:jelpf1} becomes
\begin{eqnarray} \nonumber
     \sum_{i=1}^{n} \widehat{V}_{i} &=& \sum_{i=1}^{n} n U - (n-1) n U -\sum_{k=1}^{R} \frac{1}{\widehat{p_k}}\left(\sum_{i=1}^{n}n U_k - (n-1) U_k\right) -n CM \\ \nonumber
     &=& n U - \sum_{k=1}^{R} \frac{1}{\widehat{p_k}} n U_k - n CM \\ \nonumber
     &=& n S_n.
\end{eqnarray}
Thus, we have
$$S_n =\frac{\sum_{i=1}^{n} \widehat{V}_{i}}{n}.$$
As $\widehat{p_k}$  converges in probability to $p_k$, by Slutsky's theorem $S_n$ has the same asymptotic properties as $S_n^{*}$, where
$$S^{*}_{n}=U-\frac{1}{p_k}U_k-CM.$$
 Let $g(x,y)= E \left[ |x-X_j| \left(1-\sum_{k=1}^{R}\frac{I(y=k,Y_j=k)}{\widehat{p}_k}\right)- CM|X_1 =x ,Y_1=y\right]$ and $\sigma_g^2= Var(g(x,y))$. As $S_n$ has the same asymptotic properties as $S_n^{*}$, using the asymptotic results from the theory of $U$-statistics, we have the following.

\begin{equation} \label{eq:jel_l1}
    \frac{\sqrt{n}S_n}{2 \sigma_g} \rightarrow N(0,1).
\end{equation} 
Further, denote $S= \frac{1}{n} \sum_{i=1}^{n} (\widehat{V_i})^2 $ and note that 
\begin{equation} \label{eq:jel_l2}
S = 4\sigma_g^2 + O(1). 
\end{equation}
 Given that $\lambda$ is the solution of
$$
  \frac{1}{n}\sum_{i=1}^{n}{\frac{\widehat{V}_{i}}{1+\lambda \widehat{V}_{i}}}=0,
$$
the log empirical likelihood ratio becomes 
$$-2log R( CM) = 2n \lambda (S_n) - n S \lambda^2 + O_p(1)$$
By expanding the log term using the Taylor series expansions and using the lemmas in Jing et al.  (2009), we can see that $\lambda = O_p(n^{-1/2})$. Thus, the leading terms become 
$$ -2 \log R(CM) = \frac{n S_n ^2}{S} + O_p(1).$$
In view of equation \eqref{eq:jel_l2}, we have
$$-2 \log R(CM)  \xrightarrow{d} \chi^2_1.$$
\end{proof}
Theorem \ref{thm_jelci} provides a Wilks-type result for the proposed $CM$ measure and forms the basis for constructing confidence intervals without requiring direct estimation of the asymptotic variance. Specifically, a JEL-based $100(1-\alpha)\%$ confidence interval for $CM$ is obtained by inverting the empirical likelihood ratio:
\begin{equation*}
\left\{ CM:-2\log R(CM)\leq \chi^2_{1-\alpha}(1)\right\},
\end{equation*}
where $\chi^2_{1-\alpha}(1)$ denotes the $(1-\alpha)$th percentile of a $\chi^2$ distribution with one degree of freedom.

The resulting confidence interval is therefore determined entirely by the empirical likelihood ratio and the corresponding chi-square calibration. This formulation avoids explicit estimation of the potentially complicated asymptotic variance and is consequently attractive when the categorical response has a relatively large number of categories. The finite-sample performance of the proposed JEL confidence interval is investigated through the simulation study presented in Section 6.

\section{Test for Independence}
Consider a continuous random variable $X$ with a distribution function $F$ in $\mathbb{R}$. Let $Y$ be a categorical random variable taking values $1,2,\ldots, R$ with probability mass function $p_k=P(Y=k),\,k=1,2,\cdots K$. 

Based on the observed data, we are interested in testing the null hypothesis:
  $$H_0:X \text{ and }  Y \text{ are independent}$$
 against the alternative hypothesis
 $$H_0:X \text{ and }  Y \text{ are dependent}.$$
 
It is important to note that the alternative hypothesis considered here is quite general, as $H_1$ only indicates a lack of independence and does not assume any specific form of dependence between $X$ and $Y$. $H_0$ being true is equivalent to $CM =0$. Hence, we consider the covariance measure $CM$ defined in equation \eqref{eqn_cm}, as a departure measure from the null hypothesis $H_0$ towards the alternative hypothesis $H_1$.  To carry out the test for independence we need to estimate $CM$.
As outlined in section 3, we can estimate the $CM$ using the theory of $U$-statistics, and the estimator is given as
$$ \widehat{CM}= \frac{1}{\binom{n}{2}} \sum_{i<j}\left((2 \max (X_i,X_j)-X_i-X_j) (1- \sum_{k=1}^{R} \frac{I(Y_i=k,Y_j=k)}{\widehat{p_k}}\right),$$
where $\widehat{p_k}= \frac{1}{n} \sum_{i=1}^{n} I(Y_i=k)$.

 From Corollary 1, $\sqrt{n} (\widehat{CM}-E(CM)$ follows a normal distribution with mean $0$ and standard deviation $\sigma^2$ 
$$\sigma^2 = \mathbf{a^T}\Sigma\mathbf{a}.$$

Now, let us investigate the limiting distribution of the test statistic under the null hypothesis $H_0$.

Let us denote $h(x_1,x_2,y_1,y_2) = (2 \max (x_1,x_2)-x_1-x_2) \left(1- \sum_{k=1}^{R} \frac{I(y_1=k,y_2=k)}{\widehat{p_k}}\right) $ as the kernel of $U$-statistic estimate $\widehat{CM^*}$. Let $$h_1(x_1,y_1) = E \left [(2 \max (X_1,X_2)-X_1-X_2) \left(1- \sum_{k=1}^{R} \frac{I(Y_1=k,Y_2=k)}{\widehat{p_k}}\right)|X_1=x_1,Y_1 =y_1 \right] $$
and let $\sigma_1 = Var(h_1(X_1,Y_1))$. Under the null hypothesis $H_0$, we can easily see that $\sigma_1=0$, thus making $h(x_1,x_2,y_1,y_2)$ degenerate of order $1$.  This implies that the CLT of $U$-statistics is not applicable in this case. In this case, the symmetric square integrable kernel $h(x_1,x_2,y_1,y_2)$  admits a series expansion according to  the Hilbert-Schmidt theory of integral equations.

Under the null hypothesis, when $X$ and $Y$ are independent, we can rewrite the kernel $h(x_1,x_2,y_1,y_2)$ as 
 $$ h(x_1,x_2,y_1,y_2) =h_X(x_1,x_2) h_Y(y_1,y_2).$$ Then, we can write the eigen decomposition of $h_X$ and $h_Y$, respectively, as follows. 

 $$h_X(x_1,x_2) = \sum_{m=1}^{\infty} \lambda_m \phi_m(x_1) \phi_m(x_2) $$
and
$$h_Y(y_1,y_2) = \sum_{k=1}^{R} \eta_k \psi_k(y_1) \psi_k(y_2) $$
where $\lambda_m$ and $\eta_k$ are the eigenvalues while  $\phi_m(x)$ and $\psi_k(y)$ are the corresponding orthonormal basis functions of $h_X(.)$ and $h_Y(.)$, respectively, according to the Hilbert-Schmidt theory. 
Then, using the limit theorem for degenerate $U$-statistics  ( Lee 2019), we have the following result. 
\begin{theorem}
    Let $\widehat{\CM}^*$ be the $U$- statistics associated with symmetric kernel of degree 2, $h(x_1,x_2,y_1,y_2)$ with degeneracy of  order 1 such as $E(h(x_1,x_2,y_1,y_2) =0$ under the null hypothesis. Then $n\widehat{\CM}^* \xrightarrow{d} \sum_{m=1}^{\infty} \sum_{k=1}^{R}\lambda_m\eta_k (Z_{mk}^2-1)$ where $Z_{mk}$ are i.i.d standard normal variables. 
\end{theorem}
\begin{proof}
Note 
\begin{eqnarray} \nonumber
   \widehat{ \CM}^*  &=& \widehat{U} - \sum_{k=1}^{R} \frac{\widehat{U}_k}{p_k}.
\end{eqnarray}
Observe that $\widehat{U}$, $\widehat{U}_k$, $k=1,2,\ldots,R$ are $U$-statistics with  kernel  $h'$ and $h_k$, respectively. Since $\widehat{\CM}^*$ is a linear combination of $U$-statistics, the result follows from the limit theorem for degenerate $U$-statistics  (Lee, 2019).
\end{proof}
By the weak law of large numbers, $\widehat p_k$ is a consistent estimator of $p_k$, and hence by Slutsky's theorem, the asymptotic distributions of $\widehat{\CM}$ and  $\widehat{\CM}^*$ are the same. Hence, we have the following corollary.

\begin{corollary}
    As $n \rightarrow \infty$, $n\widehat{CM} \xrightarrow{d} \sum_{m=1}^{\infty} \sum_{k=1}^{R}\lambda_m\eta_k (Z_{mk}^2-1)$ where $Z_{mk}$ are i.i.d standard normal variables. 
\end{corollary}
 For extending the testing of independence between categorical random variable $Y$ and random vector$\mathbb{X}=(X_1,X_2,\ldots,X_p)^t$ we can consider the multivariate $CM$ as given in equation (\ref{eqn_cm}) as  departure measure and the test statistics in such a case is given by 
$$ \widehat{CM}= \frac{1}{\binom{n}{2}} \sum_{i<j}||\mathbb{X}_i-\mathbb{X}_j||_1 \left(1- \sum_{k=1}^{R} \frac{I(Y_i=k,Y_j=k)}{\widehat{p_k}}\right).$$  


\section{Simulation Study}
This section presents a Monte Carlo simulation study carried out in two parts. In the first part, we evaluated the finite sample performance of the JEL-based confidence interval for the proposed covariance measure, while in the second part, we compared the finite sample performance of the proposed independence test with prevailing tests.  The simulation is conducted thousands of times using different sample sizes of 20, 40, 60, 80 and 100. The simulation is carried out using R software.

\subsection{Simulating the JEL based confidence interval for CM}
To simulate the JEL-based confidence interval for the proposed CM, we generated three distinct combinations of dependent continuous variable $X$ and discrete random variable $Y$. In each case, we calculated the confidence intervals $95\%$ of the $CM$ and reported the probability of coverage along with the average width of the estimated interval. To create a case of dependent variables $X$ and $Y$, we sampled the conditional distribution $F_{X|Y}$, which consists of mixtures of normal distributions. The three cases considered in this simulation study are  
\begin{itemize}
    \item Case $1$: $F_{X|Y} \sim \frac{1}{3}N(0,1)+\frac{1}{3}N(1,1)+\frac{1}{3}N(2,1)$
    \item Case $2$: $F_{X|Y} \sim \frac{1}{3}N(-1,1)+\frac{1}{3}N(0,1)+\frac{1}{3}N(2,1)$

    \item Case $3$: $F_{X|Y} \sim \frac{1}{3}N(-2,1)+\frac{1}{3}N(0.5,1)+\frac{1}{3}N(1.5,1)$
\end{itemize}
The results of the simulation are presented in Table \ref{tab:CI}. The results presented in Table \ref{tab:CI} indicate that the proposed approach to determine the JEL-based confidence interval for $CM$ has a satisfactory coverage probability for all scenarios considered in the simulation study. Moreover, the average width of the confidence interval is seen to decrease as the sample size increases.

\subsection{Simulation Study of Test for Independence}
We carried out the simulation study in two sub-parts. First, we find the type I empirical error considering a continuous random variable $X$ following different continuous distributions and $Y$, a categorical random variable  following different distributions. We independently generated $X$ and $Y$ to find the type I empirical error of the proposed test. In this case, we examine different scenarios by varying the assumed distribution of $X$ as a normal, exponential, and lognormal distribution while $Y$ taken as a categorical variable with $R=3$ taking values $Y \in \{1,2,3\}$ with probabilities $\{p_1,p_2,p_3\}$. For each choice of $X$, $\mathbf{p} =(p_1,p_2,p_3) $ is varied from balanced $\mathbf{p}=(1/3,1/3,1/3)$ to lightly unbalanced $\mathbf{p}=(5/12,4/12,3/12)$ to highly unbalanced $\mathbf{p}=(6/10,3/10,1/10)$. The results of the simulation study are presented in Table \ref{tab:normtyp1}, \ref{tab:exptyp1} and \ref{tab:lnormtyp1}. Additionally, we compared the empirical Type 1 error of the proposed test with some other tests: the mean-variance index test (MV) (Cui and Zhong, 2019), the integral Pearson chi-square test (IPC) (Ma et al., 2023) and the semi-distance test (SD) (Zhong et al., 2024). MV and SD tests are available in the R package `semiist'.

\begin{table}[ht]
\caption{Coverage probability and Average Width of the Confidence Intervals}
    \centering
    \begin{tabular} {cccccc}
    \hline
\hline
&\small{Parameter} &\small{$n$} & \small{Coverage Probability}  & \small{Average Width}\\
\hline
\hline
\multirow{5}{*}{ Case $1$} & \multirow{5}{*}{$0.336$} & &   & \\
&  &20 & 0.972 &1.395 \\
 &   &40 & 0.970 &0.879 \\
 &   &60 & 0.971&0.638 \\
  &  &80 & 0.963 & 0.515\\
 &   &100 & 0.958 & 0.443\\
    \hline
    \multirow{5}{*}{Case $2$} & \multirow{5}{*}{$0.696$} & &  &  \\
& &20 & 0.959 &1.673 \\
&    &40 & 0.963 &1.036 \\
  &  &60 & 0.964 &0.731\\
  &  &80 & 0.964 & 0.599\\
&    &100 & 0.957 & 0.542\\
   \hline
\multirow{5}{*}{Case $3$} & \multirow{5}{*}{$0.903$}& & & \\
&&20 & 0.962 &1.960 \\
 &   &40 & 0.971 &1.166 \\
  &  &60 & 0.965&0.892\\
 &   &80 & 0.961 & 0.793\\
 &   &100 & 0.951 & 0.606\\
    \hline
    \end{tabular}
    
    \label{tab:CI}
\end{table}

\begin{table} [H]
\caption{Empirical type I error for when $X \sim N(0,1)$}
    \centering
    \begin{tabular} {cccccc}
    \hline
\hline
\multirow{5}{*}{ Balanced} & $n$ & Proposed test  & IPC & MV & SD\\
\hline
\hline
   &30& 0.049 &0.052 &0.044 & 0.046 \\
    &60 &  0.057&0.071 &0.049 & 0.049\\
    &90 & 0.053 & 0.069&0.048 & 0.051\\
    &120& 0.042 & 0.058&0.044 &0.043\\
    &150 & 0.058 & 0.083& 0.060&0.056\\
    \hline
    \multirow{5}{*}{Lightly Balanced} &  & & & \\
 &30 & 0.056 & 0.066& 0.047& 0.053 \\
    &60 &0.057  &0.068 & 0.049&0.049 \\
    &90 &  0.046&0.058 & 0.046&0.042\\
    &120 & 0.049 &0.065 & 0.057&0.049\\
    &150 &  0.052& 0.078&0.049 &0.051\\
   \hline
\multirow{5}{*}{Highly Balanced} & &  & \\
&30 &  0.049&0.052 & 0.044&  0.046\\
    &60 & 0.057 &0.071 &0.057 &0.049 \\
    &90 & 0.053 & 0.069& 0.048& 0.051\\
    &120 & 0.043 &0.058 & 0.044&0.043\\
    &150& 0.052 &0.083 &0.066 &0.056\\
    \hline
    \end{tabular}
    
    \label{tab:normtyp1}
\end{table}

\begin{table}[H]
\caption{Empirical type I error for when $X \sim Exp(1) $}
    \centering
    \begin{tabular} {cccccc}
    \hline
\hline
\multirow{5}{*}{ Balanced} & $n$ & Proposed test  & IPC & MV & SD\\
\hline
\hline
   &30 & 0.046 &0.064 & 0.042& 0.048 \\
    &60 &  0.036&0.057 &0.042 &0.039 \\
    &90 &  0.041& 0.060&0.043 & 0.045\\
    &120 & 0.060 & 0.073& 0.056&0.060\\
    &150 & 0.049 &0.055 & 0.041&0.045\\
    \hline
    \multirow{5}{*}{Lightly Balanced} &  & & & \\
 &30 & 0.044 &0.071 &0.055 &0.055  \\
    &60 & 0.057 &0.073 & 0.057& 0.056\\
    &90 &  0.049& 0.061& 0.044&0.044\\
    &120 &0.057  & 0.087& 0.071&0.061\\
    &150 &  0.052&0.069 & 0.048&0.054\\
   \hline
\multirow{5}{*}{Highly Balanced} & &  & \\
&30 & 0.049 &0.071 &0.055 & 0.055 \\
    &60 & 0.057  &0.073 &0.057 &0.056 \\
    &90 &  0.049& 0.061& 0.044&0.044 \\
    &120 &0.057  &0.087 &0.071 &0.061\\
    &150 & 0.052 &0.069 &0.048 &0.054\\
    \hline
    \end{tabular}
    
    \label{tab:exptyp1}
\end{table}

\begin{table}[H]
\caption{Empirical type I error for when $X \sim Lognormal (0,1) $}
    \centering
    \begin{tabular} {cccccc}
    \hline
\hline
\multirow{5}{*}{ Balanced} & $n$ & Proposed test  & IPC & MV & SD\\
\hline
\hline
   &30 & 0.028 &0.061 &0.047 & 0.039 \\
    &60 &  0.039&0.059 & 0.042& 0.050\\
    &90 & 0.045 &0.065 &0.050 & 0.059 \\
    &120 & 0.054 & 0.068& 0.039&0.059\\
    &150 & 0.052 & 0.072&0.045 &0.058\\
    \hline
    \multirow{5}{*}{Lightly Balanced} &  & & & \\
  &30 & 0.028 &0.061 &0.047 & 0.039 \\
    &60 &  0.039&0.059 & 0.042& 0.050\\
    &90 & 0.045 &0.065 &0.050 & 0.059 \\
    &120 & 0.054 & 0.068& 0.039&0.059\\
    &150 & 0.052 & 0.072&0.045 &0.058\\
   \hline
\multirow{5}{*}{Highly Balanced} & &  & \\
&30 & 0.042 &0.066 &0.048 &  0.043\\
    &60 & 0.041 &0.058 & 0.049&0.047 \\
    &90 &  0.043& 0.064&0.046 & 0.047\\
    &120 &  0.051&0.059 &0.048 &0.056\\
    &150 & 0.050 & 0.06& 0.049&0.057\\
    \hline
    \end{tabular}
    
    \label{tab:lnormtyp1}
\end{table}

In the second subpart of the simulation, we evaluate the empirical power of the JEL ratio test and compare it to the MV, IPC, and SD tests. We generate $Z$ as mixtures of continuous distributions such that 
$Z \sim F_{X|Y} $

where $F_{X|Y}$ is the conditional distribution of the continuous random variable  $X$ with respect to the categorical random variable $Y$. Let $Y$

For the purpose of this study, we consider three different mixtures

We have considered different combinations of the mixture distribution. For the purpose of this simulation, we consider the following three cases.

\begin{itemize}
    \item Case 1:  $$X \sim p_1 \, \mathcal{N}(0,1) +p_2 \, Exp(1) +p_3 \, Lognormal(0,1)$$

     \item Case 2:  $$X \sim p_1 \, Exp(1) +p_2 \, Lognormal(0,1)  +p_3 \, Logistic(2,1) $$

      \item Case 3:  $$X \sim p_1 \, \mathcal{N}(0,1)   +p_2 \, Logistic(2,1) +p_3 \, t (1) $$
\end{itemize}

In addition, for each mixture distribution case considered, $ \mathbf{p} =(p_1,p_2,p_3) $ is varied from balanced $\mathbf{p}=(1/3,1/3,1/3)$ to lightly unbalanced $\mathbf{p}=(5/12,4/12,3/12)$ to highly unbalanced $\mathbf{p}=(6/10,3/10,1/10)$. Consequently, we look at the following three distinct scenarios for every scenario. For Case 1, three scenarios are presented as follows.
\begin{itemize}
    \item Scenario 1: Balanced
    $$X \sim \frac{1}{3}\, \mathcal{N}(0,1) +\frac{1}{3}\, Exp(1) +\frac{1}{3}\, Lognormal(0,1)$$

\item Scenario 2: Lightly unbalanced
$$X \sim \frac{5}{12}\, \mathcal{N}(0,1) + \frac{4}{12}\, Exp(1) + \frac{3}{12} \,Lognormal(0,1)$$

\item Scenario 3: Highly unbalanced
    $$X \sim \frac{6}{10}\, \mathcal{N}(0,1) + \frac{3}{10}\, Exp(1) +\frac{1}{10}\, Lognormal(0,1).$$

\end{itemize}
The results for  the power comparison are given in Tables \ref{tab:cas1}, \ref{tab:cas2} and \ref{tab:cas3}. From these tables,  we see that, in all scenarios, the empirical power of the test converges to one as the sample size increases. From the table, it is clear that the proposed measure has a comparable power to the existing measures for all the cases considered in this paper.

\begin{table}[H]
\caption{Empirical power at 5\% level  for Case 1.}
    \centering
    \begin{tabular} {cccccc}
    \hline
\hline
\multirow{5}{*}{ Balanced} & $n$ & Proposed test  & IPC & MV & SD\\
\hline
\hline
   &30 & 0.999 &0.999 &0.999 & 0.999 \\
    &60 &  1.000 &1.000  & 1.000 & 1.000 \\
    &90 & 1.000  &1.000 &1.000  & 1.000 \\
    &120 &1.000  & 1.000 & 1.000 &1.000 \\
    &150 & 1.000  & 1.000 &1.000  &1.000 \\
    \hline
    \multirow{5}{*}{Lightly Balanced} &  & & & \\
  &30 & 1.000  &1.000  &1.000  &1.000 \\
    &60 & 1.000 &1.000 & 1.000 & 1.000 \\
    &90 & 1.000  &1.000  &1.000  & 1.000  \\
    &120 &1.000  &1.000 & 1.000 &1.000 \\
    &150 &1.000 & 1.000 &1.000  &1.000 \\
   \hline
\multirow{5}{*}{Highly Balanced} & &  & \\
&30 & 0.989 &0.991 &0.989 &  0.995\\
    &60 &1.000 &1.000 &1.000 &1.000 \\
    &90 &1.000  & 1.000& 1.000& 1.000\\
    &120 &  1.000&1.000 &1.000 &1.000\\
    &150 &  1.000& 1.000& 1.000&1.000\\
    \hline
    \end{tabular}
        \label{tab:cas1}
\end{table}

\begin{table}[H]
\caption{Empirical power at 5\% level  for Case 2.}
    \centering
    \begin{tabular} {cccccc}
    \hline
\hline
\multirow{5}{*}{ Balanced} & $n$ & Proposed test  & IPC & MV & SD\\
\hline
\hline
   &30 & 0.446&0.453 &0.463 & 0.465 \\
    &60 &  0.684&0.701 & 0.698& 0.703\\
    &90 & 0.945&0.949&0.953 & 0.964 \\
    &120 & 1.000  & 1.000 & 1.000 &1.000 \\
    &150 & 1.000  &1.000 &1.000  &1.000 \\
    \hline
    \multirow{5}{*}{Lightly Balanced} &  & & & \\
  &30 & 0.467 &0.459 &0.498& 0.501 \\
    &60 &  0.706&0.693 & 0.713& 0.708\\
    &90 &  0.999 &0.999 & 0.999&0.999 \\
    &120 & 1.000  & 1.000 & 1.000 &1.000 \\
    &150 & 1.000  & 1.000 &1.000  &1.000 \\
   \hline
\multirow{5}{*}{Highly Balanced} & &  & \\
&30 & 0.451 &0.457 &0.445 &  0.461\\
    &60 & 0.671 & 0.681& 0.701&0.691 \\
    &90 & 0.987  &0.961 &0.989 &0.995  \\
    &120 &  1.000 &1.000  & 1.000 &1.000 \\
    &150 & 1.000 &1.000  & 1.000 &1.000 \\
    \hline
    \end{tabular}
        \label{tab:cas2}
\end{table}

\begin{table}[H]
\caption{Empirical power at 5\% level  for Case 3.}
    \centering
    \begin{tabular} {cccccc}
    \hline
\hline
\multirow{5}{*}{ Balanced} & $n$ & Proposed test  & IPC & MV & SD\\
\hline
\hline
   &30 & 0.761 &0.920 &0.880 & 0.900 \\
    &60 &  0.871&0.951 & 0.942& 0.965\\
    &90 & 0.991 &0.990 &0.999 & 1.000 \\
    &120 & 1.000 & 1.000& 1.000&1.000\\
    &150 & 1.000 & 1.000&1.000&1.000\\
    \hline
    \multirow{5}{*}{Lightly Balanced} &  & & & \\
  &30 & 0.804 &0.901 &0.854 & 0.915 \\
    &60 &  0.961&0.967 & 0.971& 0.989\\
    &90 & 1.000&1.000 &1.000& 1.000\\
    &120 & 1.000 & 1.000& 1.000&1.000\\
    &150 & 1.000 & 1.000&1.000 &1.000\\
   \hline
\multirow{5}{*}{Highly Balanced} & &  & \\
&30 & 0.901 &0.906 &0.961  &  0.965\\
    &60 & 0.984 & 0.998& 0.999& 0.999\\
    &90 &  1.000& 1.000&1.000 &1.000 \\
    &120 & 1.000 & 1.000& 1.000&1.000\\
    &150 &  1.000&1.000 & 1.000&1.000\\
    \hline
    \end{tabular}
        \label{tab:cas3}
\end{table}

\section{Data Analysis}

\begin{figure}[H]
\centering
\caption{Comparison of Flower Species across different variables}
\vspace{0.2in}
\begin{tabular}{c c }
 \includegraphics[width=7cm]{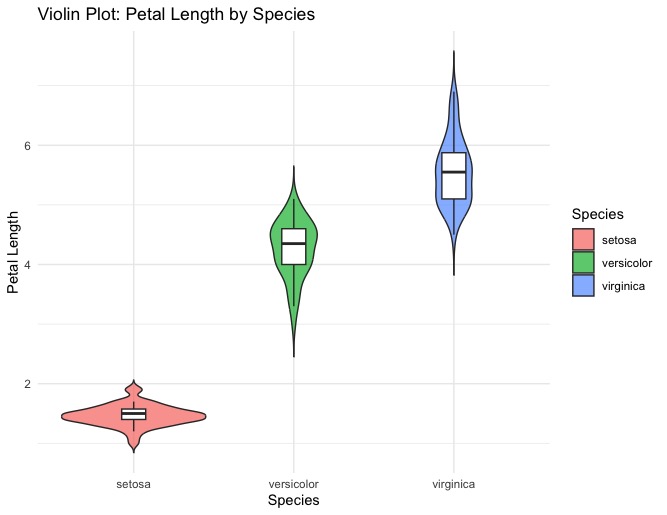}   
    &\includegraphics[width=7cm]{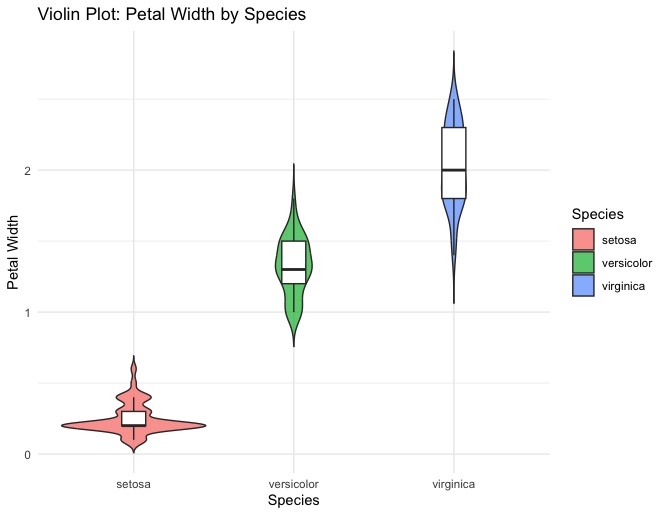}\\
  \includegraphics[width=7cm]{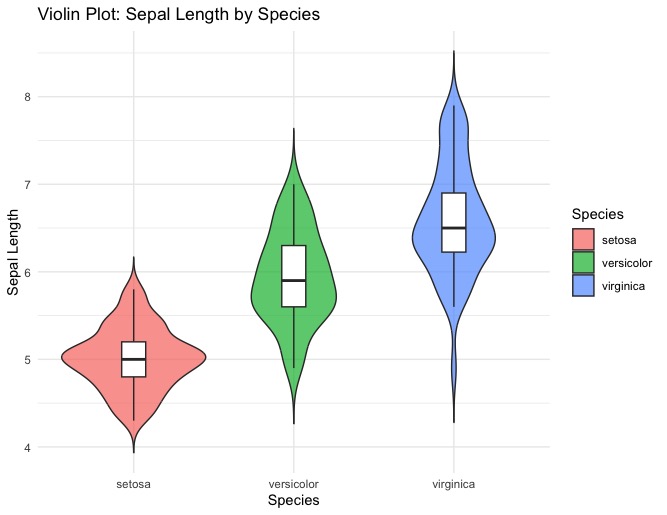}   
    &\includegraphics[width=7cm]{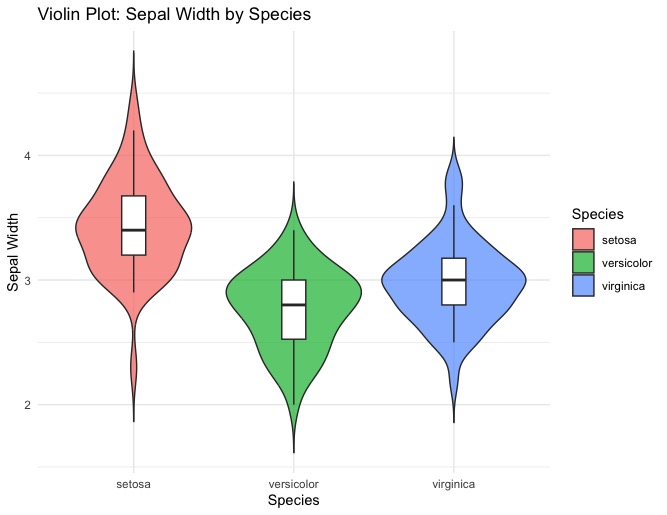}  
\end{tabular}

\label{fig:vio_flo}

\end{figure}

In this section, we demonstrate the application of the proposed measure $CM$ in two distinct data sets. The Iris dataset, originally introduced by British statistician and biologist Ronald A. Fisher in 1936, is the first dataset considered in this study. The dataset comprises 150 observations of iris flowers sampled from three distinct species: Iris setosa, Iris versicolor, and Iris virginica. Each observation includes four distinct morphological measurements: sepal length, sepal width, petal length, and petal width. The three species display distinct patterns of morphological variation, rendering this dataset appropriate for illustrating the proposed measure for variable selection in species classification. Figure \ref{fig:vio_flo} illustrates the variations in four distinct morphological features among these species. 

To demonstrate the application of $CM$, we have computed the $CM$ values to examine the relationship between the Iris species variable and various subsets of four morphological characteristics. Table \ref{tab:sss_iris} presents the $CM$ values for different combinations of the morphological variables.

\begin{table}[H]
\centering
\caption{SSS-based Variable Selection Results for the Iris Dataset}
\label{tab:sss_iris}
\renewcommand{\arraystretch}{1.2}
\begin{tabular}{l S[table-format=1.3] S[table-format=1.3] S[table-format=1.3]}
\hline
\textbf{Variable Subset} & \textbf{CM} & \textbf{CI\_L} & \textbf{CI\_U} \\
\hline
Full Model (All 4 Variables)                          & 6.970 & 6.570 & 7.410 \\
Sepal.Length + Sepal.Width + Petal.Length             & 5.446 & 5.086 & 5.805 \\
Sepal.Length + Sepal.Width + Petal.Width              & 3.455 & 3.241 & 3.668 \\
Sepal.Length + Petal.Length + Petal.Width             & 6.369 & 5.933 & 6.805 \\
Sepal.Width + Petal.Length + Petal.Width              & 5.639 & 5.292 & 5.987 \\
Sepal.Length + Petal.Length                           & 4.845 & 4.491 & 5.198 \\
Sepal.Length + Petal.Width                            & 2.854 & 2.638 & 3.069 \\
Sepal.Width + Petal.Length                            & 4.115 & 3.850 & 4.381 \\
Sepal.Width + Petal.Width                             & 2.125 & 1.991 & 2.258 \\
Sepal.Length + Sepal.Width                            & 1.931 & 1.789 & 2.073 \\
Petal.Length + Petal.Width                            & 5.039 & 4.707 & 5.370 \\
Petal.Length                                          & 3.514 & 3.271 & 3.757 \\
Petal.Width                                           & 1.524 & 1.418 & 1.629 \\
Sepal.Length                                          & 3.514 & 3.271 & 3.757 \\
Sepal.Width                                           & 0.601 & 0.518 & 0.683 \\
\hline
\end{tabular}
\end{table}

The highest value of $CM$ is $6.970$, when all four variables are used. This can be interpreted as if the combination of all four variables helps in achieving the maximum separation of the three Iris species. Among the 3 variable subsets, Sepal. Length + Petal. Length + Petal. Width is the best-performing combination, which is quite close to the full model. To further our understanding, we have constructed different multinomial logit models with all four variables as well as the variations with all three-variable combinations. The McFadden $R^2$ value of each model is calculated and presented in Table \ref{tab:logit_compare}.
\begin{table}[H]
\centering
\caption{Comparison of Multinomial Logistic Regression Models on the Iris Dataset}
\label{tab:logit_compare}
\renewcommand{\arraystretch}{1.2}
\begin{tabular}{lccc}
\toprule
\textbf{Model} & \textbf{McFadden's $R^2$} \\
\midrule
Full Model (4 variables)                & 0.963 \\
Sepal.Length + Sepal.Width + Petal.Length & 0.927\\
Sepal.Length + Sepal.Width + Petal.Width  &0.921\\
Sepal.Length +Petal.Length + Petal.Width  & 0.952 \\
Sepal.Width + Petal.Length + Petal.Width&0.959\\
\bottomrule
\end{tabular}
\end{table}

Table \ref{tab:logit_compare} indicates that the full four-variable model demonstrates the highest performance, as confirmed by a $R^2$ value of $0.963$. All three variable models exhibit high explanatory power, with $R^2$ values ranging from $0.921$ to $0.959$. It is worth noting that the combination of Sepal.Width, Petal.Length, and Petal.Width yields a McFadden's $R^2$ of $0.959$, which is marginally higher than the $0.952$ obtained from the model incorporating Sepal.Length, Petal.Length, and Petal.Width. Nevertheless, class separability analysis based on $CM$ indicates that the latter combination is better. While $CM$ is a non-parametric, $L_1$ distance-based measure that captures the geometric separability of classes, $R^2$ evaluates the statistical fit on how well the predictors are linearly related to the log -odds ratio of the class memberships. The combination of Petal.Length and Petal.Width exhibits a high CM score, signifying its significant contribution to class separability. Consequently, the inclusion of the Petal.Length and Petal.Width pair in the model results in only marginal differences when a third variable, either Sepal.Length or Sepal.Width, is added to the logistic model. Sepal.Width exhibits a more linear pattern in class separation, as illustrated in Figure \ref{fig:vio_flo}, which may have resulted in a marginally higher McFadden's $R^2$. However, Sepal.Length offers a stronger $L_1$ geometric separation. 

The biomarker data from the Alzheimer's Disease Neuro-imaging Initiative (ADNI), a large longitudinal research project that was started in 2004, is the second dataset used in the study. The primary goal of ADNI is to find and validate biomarkers that can track the development of mild cognitive impairment (MCI), Alzheimer's disease (ALD), and normal aging in order to aid in the development of early diagnostic tools and therapeutic monitoring strategies.

The dataset employed in this study includes baseline information on demographic characteristics such as age, sex, education, and clinical diagnosis. It also encompasses global cognition variables, including MMSE, CDR-SB, and ADAS-Cog, as well as episodic memory test scores from RAVLT and Logical Memory. Additionally, it contains measures of executive function, such as the Trail Making Test Part B and Digit Symbol, functional status assessed by the FAQ, and composite cognitive scores represented by mPACC. The dataset comprises biological markers from cerebrospinal fluid (CSF), including A$\beta$42, total tau (t-tau), and phosphorylated tau (p-tau), as well as genetic risk information, specifically APOE-$\varepsilon 4$ alleles.
 
The present study chose 19 different variables from the complete dataset to demonstrate the application of $CM$ in identifying variables that contribute to explaining the Alzheimer Diagnosis at baseline ($DX_bl$). The selected variables are detailed in Table \ref{tab:adni_des}.
\begin{table}[H]
\centering
\scriptsize
\caption{List of Variables Included in the Analysis of ADNI Dataset}
\label{tab:adni_des}
\begin{tabular}{llp{9cm}}
\toprule
\textbf{Variable} & \textbf{Category} & \textbf{Description} \\
\midrule
ABETA\_bl               & CSF Biomarker & CSF A$\beta$42 concentration; A biomarker whose reduced levels reflect amyloid plaque deposition. \\
TAU\_bl                 & CSF Biomarker & Total tau; A biomarkr whose elevated levels indicate neuronal degeneration. \\
PTAU\_bl                & CSF Biomarker & Phosphorylated tau; A biomarker of tau tangle pathology. \\
MMSE\_bl                & Global Cognition & Mini-Mental State Examination;  It is awidely used cognitive screening tool. \\
CDRSB\_bl               & Clinical Dementia Rating & CDR Sum of Boxes; A functional and cognitive impairment measure. \\
ADAS11\_bl              & Cognitive Function & ADAS-Cog 11; This evaluates memory, language, and praxis. \\
ADAS13\_bl              & Cognitive Function & ADAS-Cog 13; An extended cognitive impairment assessment. \\
ADASQ4\_bl              & Memory & ADAS Question 4 (delayed word recall). \\
RAVLT\_immediate\_bl    & Memory & A memory test:  Rey Auditory Verbal Learning Test: immediate recall total. \\
RAVLT\_learning\_bl     & Memory & A memory test: RAVLT learning slope (Trial 5 minus Trial 1). \\
RAVLT\_forgetting\_bl   & Memory &  A memory test:RAVLT forgetting score (immediate minus delayed). \\
RAVLT\_perc\_forgetting\_bl & Memory &  A memory test:Percent forgetting on RAVLT. \\
LDELTOTAL\_BL           & Logical Memory & Logical memory delayed recall score. \\
DIGITSCOR\_bl           & Attention/Processing & Digit Symbol Substitution Test score. \\
TRABSCOR\_bl            & Executive Function & Trail Making Test Part B time; executive functioning. \\
FAQ\_bl                 & Functional Ability & Functional Activities Questionnaire; daily functioning. \\
mPACCdigit\_bl          & Composite Score & Memory + Digit Symbol-based PACC composite. \\
mPACCtrailsB\_bl        & Composite Score & Memory + Trails-B PACC composite. \\
APOE4                   & Genetic Marker & Number of APOE-$\varepsilon 4$ alleles;  A genetic risk factor for AD. \\
\bottomrule
\end{tabular}
\end{table}

For each selected variable, we calculated the $CM$ and the Semi-distance Covariance measure $SDCov$ in relation to the categorical variable $DX_{bl}$, which represents the baseline diagnosis of the individual. The calculated values for each variable are presented in Table \ref{tab:var_imp}.

\begin{table}[H]
\centering
\scriptsize
\caption{Variable Importance Ranking Using CM and Semi-Distance Covariance Measure}
\label{tab:var_imp}
\begin{tabular}{clccp{13cm}}
\toprule
\textbf{Rank} & \textbf{Variable} & \textbf{CM Value} & \textbf{SDCov} \ \\
\midrule
1  & ABETA\_bl                  & 449.950 & 45.240\\                 
2  & TAU\_bl                    & 154.060 & 14.828 \\
3  & TRABSCOR\_bl               & 104.010&  13.160 \\
4  & RAVLT\_perc\_forgetting\_bl & 47.570 &  8.734\\
5  & PTAU\_bl                   & 17.630 &   3.267\\
6  & DIGITSCOR\_bl              & 16.410 & 3.189 \\
7  & RAVLT\_immediate\_bl       & 15.490&   3.142\\
8  & ADAS13\_bl                 & 13.310 & 3.141 \\
9  & mPACCtrailsB\_bl           & 9.550  & 2.719\\                       
10 & mPACCdigit\_bl             & 9.480 & 2.336\\
11 & FAQ\_bl                    & 8.920 & 2.236 \\
12 & ADAS11\_bl                 & 8.430 & 1.711\\                      
13 & LDELTOTAL\_BL              & 8.400 & 1.556   \\
14 & ADASQ4\_bl                 & 4.270 & 1.120\\
15 & MMSE\_bl                   & 4.020 &1.070\\ 
16 & RAVLT\_learning\_bl        & 3.630 & 1.050 \\                       
17 & CDRSB\_bl                  & 3.040 & 0.563  \\
18 & RAVLT\_forgetting\_bl      & 2.750& 0.099 \\
19 & APOE4                      & 0.800& 0.073 \\
\bottomrule
\end{tabular}
\end{table}

Table \ref{tab:var_imp} indicates that the four variables ABETA\_bl, TAU\_bl, TRABSCOR\_bl, and  RAVLT\_perc\_forgetting\_bl exhibit the highest values of $CM$ and $SDCov$, suggesting that these variables may be the most correlated with $DX_{bl}$ relative to the others. Note that these highly ranked variables are the core AD biomarkers and cognitive markers which is coinciding with scientific literature.  Additionally, both $CM$ and $SDCov$ exhibit comparable rankings for weaker related variables. The weakest variable APOE4 serves as a risk factor rather than a diagnostic marker, aligning the results with existing scientific literature. It is evident that $CM$ offers a more pronounced distinct
ion between important and trivial variables in comparison to $SDCov$. When feature selection is the objective, $CM$ offers a more precise ranking by distinctly differentiating between significant and trivial variables. The values of $SDCov$ are smaller and exhibit more smoothness, thereby obscuring the differences among variables.

In the following Figure \ref{fig:adnielbow}, we present the elbow plots of the ranked variable based on the $CM$ and $SDCov$ measures, along with a log-scale comparison of the two. The $CM$-based elbow plot exhibits a clear and steep decline in values from rank 1 to 4, after which the decay is smoother. The sharp change in the slope suggesting these four top variables might capture the dependence structure and contribute statistically. 

In $SDCov$ based elbow plot, we can seen that curve is decaying more smoothly in smaller magnitudes. While we can identify the elbow in this plot, it is not discernible as the $CM$ plot. 

Further, from the log-scale comparison plot, we can see two curves are approximately parallel across all ranks of variables, showing a consistency in the relative ordering of the variables. However, the sharper elbow behavior in the $CM$ provides better visual guidance for variable selection, highlighting the measure's utility in detecting relevant predictors in dimension reduction problems.

\begin{figure}[H]
\centering
\caption{Comparison of the Elbow plots for the CM and SDCov value calculated }
\vspace{0.2in}
\begin{tabular}{c c}
 \includegraphics[width=7cm]{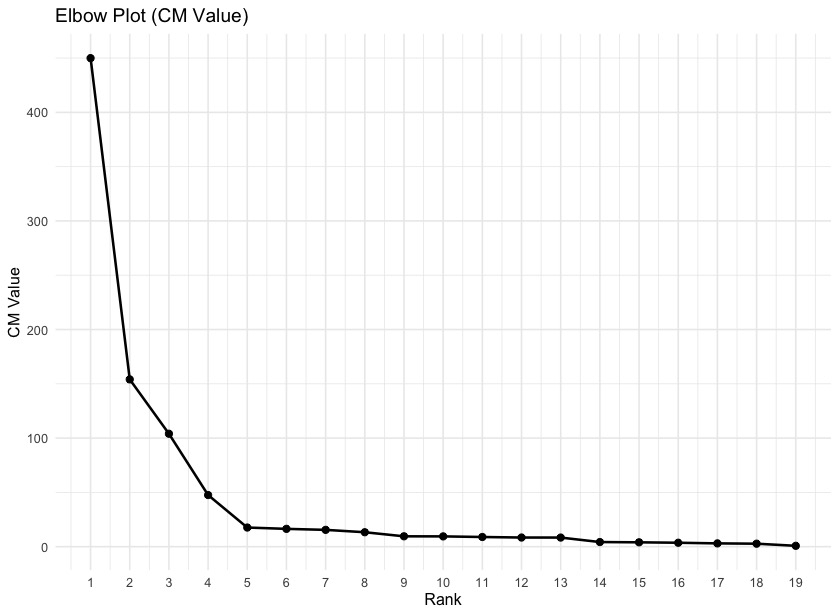}   
 &\includegraphics[width=7cm]{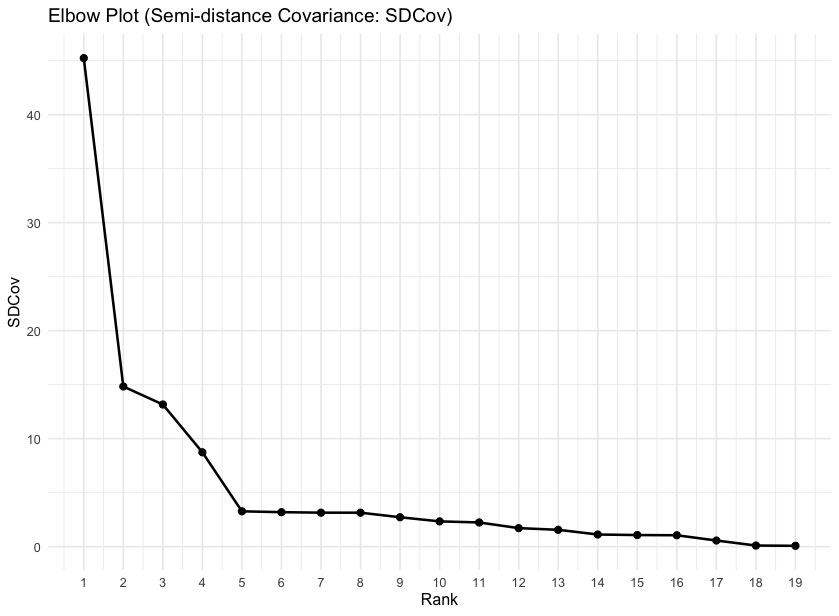}\\
\end{tabular}
\includegraphics[width=7cm]{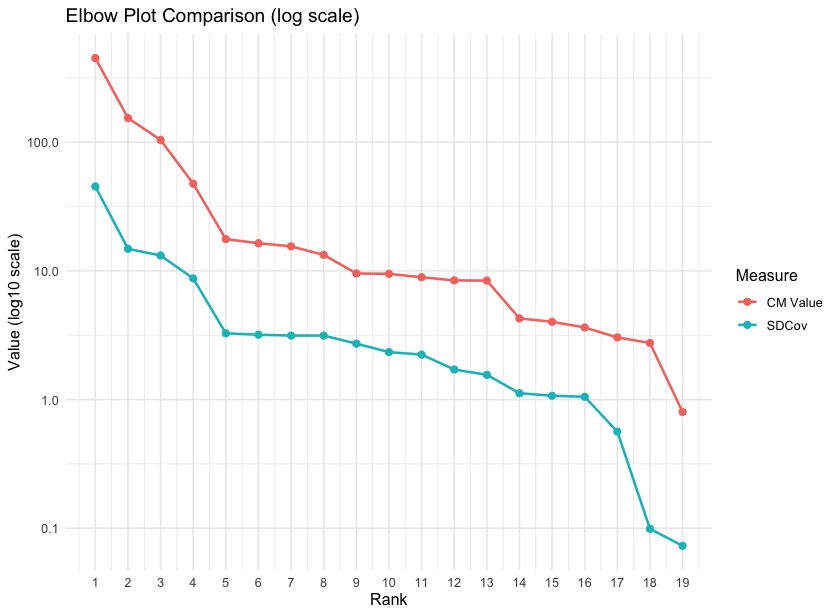}
\label{fig:adnielbow}
\end{figure}

\begin{figure}[H]
\centering
\caption{Comparison of consecutive log-ratio: CM and SDCOv }
\vspace{0.2in}
\
\includegraphics[width=12 cm]{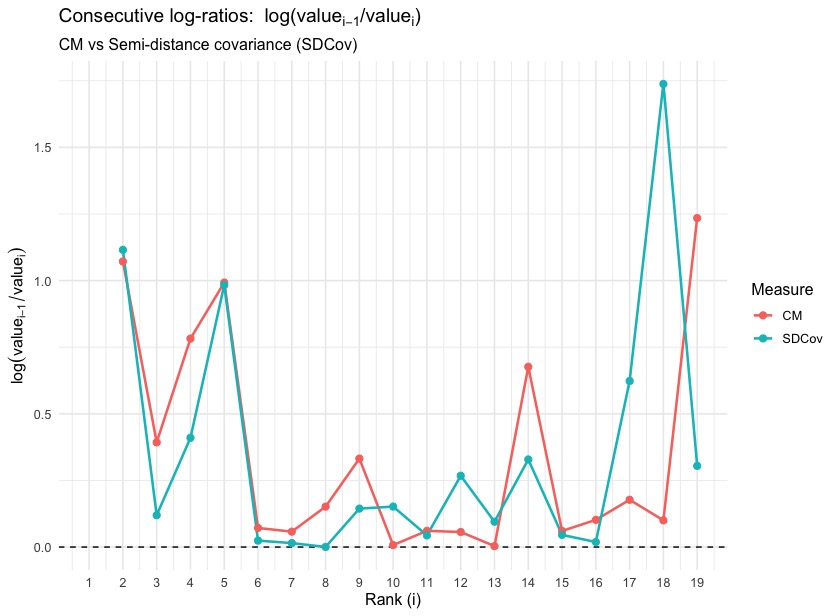}
\label{fig:log_adni}
\end{figure}

In Figure \ref{fig:log_adni}, we have compared the consecutive log ratio of the $CM$ and $SDCov$ measures. This plot reveals that early ranks, we get pronounced spikes showing that there is significant drop in dependence strength after top ranked variables. After the rank 6 variable, we can see that values of the ratios are very low, indicating that there is little change in dependency measure indicating that these lower rank variables might contribute only marginally for explaining the dependence.

Further, from Figure \ref{fig:log_adni}, we can also note that $CM$ provides clear structural break in identifying the important variables. Note that for $CM$ curve, there is a high volatility among the top ranked variables while the in the late ranked variables, the log ratios are more stable with values closer to 0. However, $SDCov$ exhibits are contrasting behavior by showing  high volatility in late variables also, that contribute to confusion in variable identification.

The use of $L_1$ norm in $CM$ makes it sensitive to a shifts in central tendency and hence more effective in measuring the class separation. While $SDCov$ uses the $L_2$ norm, which is effective in capturing the dependency between $X$ and $Y$ and it is more sensitive to variance differences. Thus, though results both measures coincides, it is more preferable to use $CM$ than $SDCov$ when the objective is variable selection especially in cases where outliers are present. 

\begin{figure}[H]
\centering
\caption{Comparison of Alzheimer's Disease diagnosis across different factors}
\vspace{0.2in}
\begin{tabular}{c c }
 \includegraphics[width=7cm]{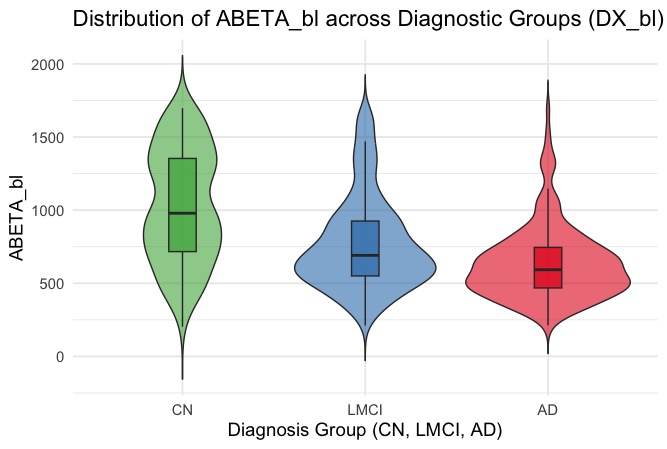}   
    &\includegraphics[width=7cm]{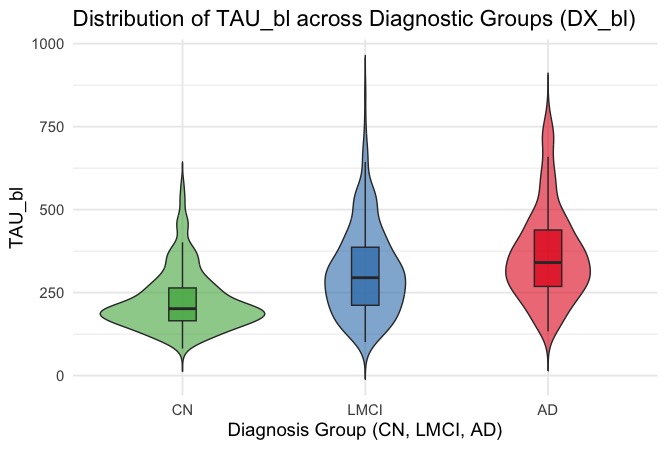}\\
  \includegraphics[width=7cm]{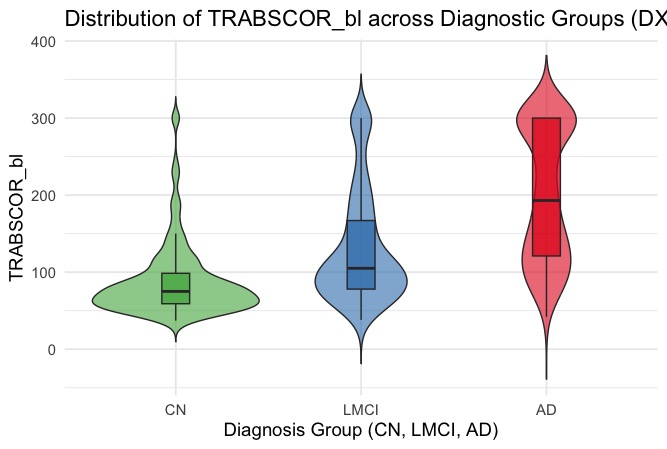}   
    &\includegraphics[width=7cm]{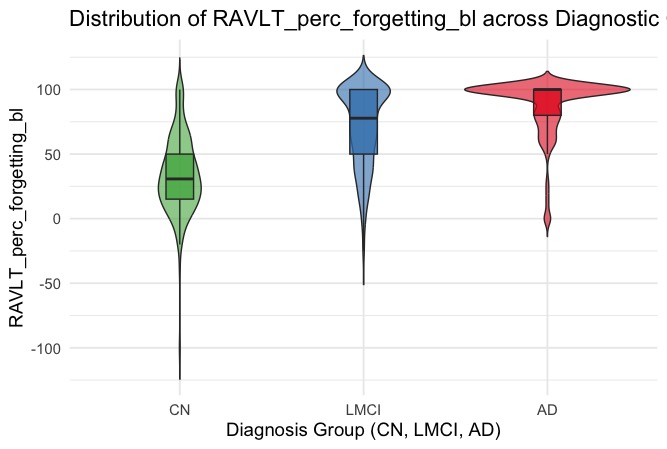}  
\end{tabular}

\label{fig:vio_AD}
\end{figure}

Now, to understand the difference among these classes of Alzheimer's Disease diagnosis with respect to these four significant variables, we have plotted the violin plots given in Figure \ref{fig:vio_AD}.

To further our understanding of the extent to which the selected variables account for diagnostic variation in the dataset, we fitted a multinomial logistic regression model using only the four highest-ranked variables based on $CM$ value. The resulting model resulted in a McFadden's $R^2$ value of 0.7407, indicating a high level of explanatory power for this type of model. This strong performance provides additional empirical support for the effectiveness of CM-based feature selection. 

\section{Conclusion and future work}

This study introduced a new dependence measure for assessing the association between a categorical random variable and a continuous random variable by modifying the semi-distance covariance framework of Zhong et al. (2024). The principal methodological modification is the replacement of the Euclidean ($L_2$) distance in the semi-distance covariance construction with the Manhattan ($L_1$) distance. This seemingly simple modification leads to a dependence measure that retains the fundamental independence characterization of the original semi-distance covariance while providing a more robust and interpretable measure of association. In particular, the use of the $L_1$ distance reduces the influence of extreme pairwise distances and is therefore particularly appealing in applications involving heavy-tailed distributions and potential outlying observations.

The theoretical development establishes the fundamental properties of the proposed measure, including its characterization of independence between the categorical and continuous variables. These properties provide the theoretical foundation for using the proposed measure as a model-free tool for dependence assessment and feature screening. Importantly, the construction does not require specification of a parametric relationship between the response and predictor variables, making the methodology applicable to a broad range of nonlinear and non-Gaussian settings. The proposed measure can also be naturally extended from the univariate setting to multivariate continuous predictors, which is particularly relevant for high-dimensional applications in which groups or combinations of variables may jointly contain information about a categorical response.

We also developed estimation and inferential procedures for the proposed measure. In particular, a $U$-statistics-based estimator was introduced and its asymptotic properties were established. To facilitate practical inference, we developed a jackknife empirical likelihood (JEL) procedure for constructing confidence intervals for the proposed dependence measure. The JEL formulation is attractive because it avoids direct estimation of the potentially complicated asymptotic variance and provides a convenient Wilks-type calibration for inference. In addition, the proposed framework leads naturally to a test of independence between continuous and categorical random variables. These inferential developments complement the descriptive role of the dependence measure and provide a systematic framework for both quantifying and testing dependence.

The extensive simulation studies demonstrate that the proposed $L_1$-based measure performs competitively with existing dependence measures across a range of data-generating mechanisms. In particular, the simulations illustrate the usefulness of the proposed measure in settings characterized by non-Gaussian behavior, heavy tails, and contamination by outlying observations. These findings support the motivation for replacing the Euclidean distance with the Manhattan distance and suggest that the proposed measure can provide a useful alternative to existing measures when robustness is an important consideration. The simulation results also demonstrate its potential as a screening criterion for identifying variables that carry information about a categorical response.

The empirical analyses further illustrate the practical utility of the proposed methodology in both simple and high-dimensional settings. The analysis of the classical Iris dataset provides an interpretable illustration of the ability of the proposed measure to identify variables associated with class separation. More importantly, its application to the Alzheimer's Disease Neuroimaging Initiative (ADNI) dataset demonstrates the potential of the methodology in a substantially more complex scientific setting involving high-dimensional predictors. The variables and variable combinations identified by the proposed measure were subsequently evaluated through multinomial logistic regression, with the resulting McFadden's $R^2$ values providing additional evidence of their usefulness for distinguishing among response categories. Together, these analyses demonstrate that the proposed measure can serve as a computationally tractable and interpretable screening criterion for identifying individual variables or groups of variables that are informative about a categorical response.

Several directions remain open for future research. First, the present development focuses primarily on the $L_1$ distance. A natural extension would be to consider a broader class of $L_q$ or other robust distance functions and investigate how the choice of distance affects the sensitivity of the resulting dependence measure to tail behavior, contamination, sparsity, and different forms of dependence. Such a framework could potentially provide a family of distance-based measures whose robustness and efficiency can be tailored to the characteristics of a particular application.

Second, the proposed methodology can be further developed for genuinely high-dimensional regimes in which the number of continuous predictors is substantially larger than the sample size. In such settings, computational scalability and the control of false discoveries become important considerations. Developing theoretical screening properties, such as the sure screening property, ranking consistency, and screening consistency under diverging dimensionality, would provide a stronger theoretical foundation for the use of the proposed measure in modern high-dimensional applications. It would also be of interest to investigate iterative or conditional screening procedures that can identify variables whose effects may be masked by marginal screening.

Third, although the present work considers a categorical response, extensions to more complex response structures are possible. For example, the framework could be adapted to ordinal or multinomial responses with a large number of categories, as well as to settings involving longitudinal, functional, or dependent observations. Developing versions of the proposed measure that accommodate structured dependence among observations would substantially broaden its applicability.

Fourth, the robustness properties of the proposed measure merit further theoretical investigation. The use of the $L_1$ distance provides an intuitive robustness advantage, but a formal investigation based on influence functions, breakdown properties, contamination models, and robustness-efficiency tradeoffs would provide a more complete understanding of its behavior. Such an analysis could also clarify the extent to which the proposed measure gains robustness at the expense of statistical efficiency under light-tailed distributions.

Finally, the computational aspects of the methodology warrant further study. Although the proposed measure is computationally tractable, evaluating pairwise distances can become expensive when both the sample size and predictor dimension are large. Efficient algorithms, approximation techniques, and scalable implementations would therefore be valuable for applications involving very large datasets. Further integration with modern variable-selection and machine-learning procedures could also provide promising avenues for developing hybrid screening frameworks that combine the robustness and interpretability of the proposed dependence measure with more sophisticated predictive modeling techniques.

Overall, this work contributes a robust, model-free, and interpretable framework for measuring and testing dependence between categorical and continuous variables. By replacing the Euclidean distance with the Manhattan distance, the proposed methodology provides a useful alternative to existing semi-distance covariance-based approaches, particularly in the presence of heavy-tailed observations and outliers. The theoretical results, inferential procedures, simulation evidence, and applications to both the Iris and ADNI datasets collectively demonstrate its potential as a practical tool for dependence assessment and feature screening. The proposed framework thus provides a foundation for further developments in robust dependence measures and high-dimensional variable screening for categorical outcomes.

\newpage
\section*{Appendix: Additional Derivations}
\begin{lemma}
Let $Y_1, Y_2$ be i.i.d.\ copies of a discrete random variable $Y$ taking values in $\{1,\ldots,R\}$ with probabilities $p_k = \Pr(Y = k) > 0$. Define
\[
V = 1 - \sum_{k=1}^R \frac{\mathbf{1}\{Y_1 = k, Y_2 = k\}}{p_k}.
\]
Then
\[
\mathbb{E}[V] = 0,
\qquad
\operatorname{Var}(V) = R - 1.
\]
\end{lemma}

\begin{proof}
We first compute the expectation. By linearity of expectation,
\[
\mathbb{E}[V]
=
1 - \sum_{k=1}^R \frac{\mathbb{E}\big[\mathbf{1}\{Y_1 = k, Y_2 = k\}\big]}{p_k}.
\]
Since $Y_1$ and $Y_2$ are independent,
\[
\mathbb{E}\big[\mathbf{1}\{Y_1 = k, Y_2 = k\}\big]
=
\Pr(Y_1 = k, Y_2 = k)
=
p_k^2.
\]
Thus,
\[
\mathbb{E}[V]
=
1 - \sum_{k=1}^R \frac{p_k^2}{p_k}
=
1 - \sum_{k=1}^R p_k
=
1 - 1
=
0.
\]

Next, define
\[
S := \sum_{k=1}^R \frac{\mathbf{1}\{Y_1 = k, Y_2 = k\}}{p_k},
\quad \text{so that} \quad V = 1 - S.
\]
Then
\[
\operatorname{Var}(V) = \operatorname{Var}(S).
\]

Write $S = \sum_{k=1}^R A_k$, where
\[
A_k := \frac{\mathbf{1}\{Y_1 = k, Y_2 = k\}}{p_k}.
\]
Note that the events $\{Y_1 = k, Y_2 = k\}$ are mutually exclusive across $k$, and hence $A_k A_\ell = 0$ for $k \neq \ell$. Therefore,
\[
S^2 = \sum_{k=1}^R A_k^2.
\]
Since
\[
A_k^2 = \frac{\mathbf{1}\{Y_1 = k, Y_2 = k\}}{p_k^2},
\]
we obtain
\[
\mathbb{E}[S^2]
=
\sum_{k=1}^R \frac{\Pr(Y_1 = k, Y_2 = k)}{p_k^2}
=
\sum_{k=1}^R \frac{p_k^2}{p_k^2}
=
\sum_{k=1}^R 1
=
R.
\]
Finally, since $\mathbb{E}[S] = 1$, we have
\[
\operatorname{Var}(S)
=
\mathbb{E}[S^2] - \big(\mathbb{E}[S]\big)^2
=
R - 1.
\]
This completes the proof.
\end{proof}
\end{document}